\documentclass[twocolumn]{aastex631}
\usepackage{amsmath}
\usepackage{multirow}
\usepackage{enumitem}
\setlist[itemize]{noitemsep} 
\newcommand{\e}[2]{$\rm #1 \times 10^{#2}$}

\newcommand{\msun}{$\rm M_{\sun}$}
\newcommand{\rsun}{$\rm R_{\sun}$}
\newcommand{\halpha}{$\rm H_{\alpha}$}
\newcommand{\hbeta}{$\rm H_{\beta}$}

\newcommand{\ergpsqcmps}{$\rm erg \ cm^{-2} \ s^{-1}$}
\newcommand{\ergpsqcmpspang}{$\rm erg \ cm^{-2} \ s^{-1} \ \text{\AA}^{-1}$}

\received{21 July, 2024}
\revised{11 November, 2024}
\accepted{16 November, 2024}

\shorttitle{MODELS OF INTERACTION--POWERED SN SPECTRA}
\shortauthors{Wagle, Chatzopoulos \& Baer}
\begin{document}

\title{Spectroscopic Modeling of Luminous Transients Powered by H-Rich and He-Rich Circumstellar Interaction}

\correspondingauthor{Gururaj A. Wagle}
\email{gururaj.wagle@ulb.be}

\author[0000-0002-3356-5855]{Gururaj A. Wagle}
\affiliation{Institut d'Astronomie et d'Astrophysique, CP-226, Universit\'{e} Libre de Bruxelles, B-1050, Brussels, Belgium}

\author[0000-0002-8179-1654]{Emmanouil Chatzopoulos}
\affiliation{Department of Physics and Astronomy, Louisiana State University, Baton Rouge, LA 70803, USA}
\affiliation{Institute of Astrophysics, Foundation for Research and Technology-Hellas (FORTH), Heraklion, 70013, Greece}

\author[0009-0004-3209-685X]{Michael J. Baer}
\affiliation{Department of Physics and Astronomy, Louisiana State University, Baton Rouge, LA 70803, USA}

%% Note that the \and command from previous versions of AASTeX is now
%% depreciated in this version as it is no longer necessary. AASTeX 
%% automatically takes care of all commas and "and"s between authors names.

%% AASTeX 6.31 has the new \collaboration and \nocollaboration commands to
%% provide the collaboration status of a group of authors. These commands 
%% can be used either before or after the list of corresponding authors. The
%% argument for \collaboration is the collaboration identifier. Authors are
%% encouraged to surround collaboration identifiers with ()s. The 
%% \nocollaboration command takes no argument and exists to indicate that
%% the nearby authors are not part of surrounding collaborations.

%% Mark off the abstract in the ``abstract'' environment. 
\begin{abstract}

In this study, we perform detailed spectroscopic modeling to analyze the interaction of circumstellar material (CSM) with ejecta in both hydrogen-rich and hydrogen-poor superluminous supernovae (SLSNe), systematically varying properties such as CSM density, composition, and geometry to explore their effects on spectral lines and light curve evolution. Using advanced radiative transfer simulations with the new, open–source SuperLite code to generate synthetic spectra, we identify key spectroscopic indicators of CSM characteristics. Our findings demonstrate that spectral lines of hydrogen and helium exhibit significant variations due to differences in CSM mass and composition. In hydrogen-rich SLSN-II, we observe pronounced hydrogen emission lines that correlate strongly with dense, extended CSM, suggesting massive, eruptive mass–loss histories. Conversely, in hydrogen-poor SLSNe, we recover mostly featureless spectra at early times, with weak hydrogen lines appearing only in the very early phases of the explosion, highlighting the quick ionization of traces of hydrogen present in the CSM. We analyze the properties of the resulting emission lines, particularly \halpha \ and \hbeta , for our models using sophisticated statistical methods. This analysis reveals how variations in the supernova progenitor and CSM properties can lead to distinct spectroscopic evolutions over time. These temporal changes provide crucial insights into the underlying physics driving the explosion and the subsequent interaction with the CSM. By linking these spectroscopic observations to the initial properties of the progenitor and its surrounding material, our study offers a useful tool for probing the pre–explosion history of these explosive events.

\end{abstract}

%% Keywords should appear after the \end{abstract} command. 
%% The AAS Journals now uses Unified Astronomy Thesaurus concepts:
%% https://astrothesaurus.org
%% You will be asked to selected these concepts during the submission process
%% but this old "keyword" functionality is maintained in case authors want
%% to include these concepts in their preprints.

%% We recommend that authors also use the natbib \citep
%% and \citet commands to identify citations.  The citations are
%% tied to the reference list via symbolic KEYs. The KEY corresponds
%% to the KEY in the \bibitem in the reference list below. 

\section{Introduction} \label{sec:intro}

Over the last twenty years, the time--domain astrophysics community has been captivated by the discovery of extremely luminous and exotic transient phenomena, sparkling rigorous theoretical studies aiming to explain their origins. The high-cadence, wider field-of-view surveys conducted by Pan-STARRS \citep[Panoramic Survey Telescope and Rapid Response System][]{Kaiser:2002aa}, the Large Synoptic Survey Telescope \citep[LSST][]{Ivezic:2008aa}, ASAS-SN \citep[All-Sky Automated Survey for Supernovae][]{Shappee:2014aa}, and ZTF \citep[Zwicky Transient Facility][]{Bellm:2019aa} are continuously discovering new instances of rare, luminous events. Among these revelations, the super luminous supernovae (SLSNe) stand out as particularly enigmatic \citep{Quimby:2011tc,Gal-Yam:2012uw}. The SLSNe, coined for their peak luminosities exceeding their classical counterparts by over 2 magnitudes, challenge our traditional understanding of the mechanisms that source the normal SNe, viz., SN shock waves and radioactive decay of $^{56}$Ni. The high luminosity of the SLSNe  make them detectable from large distances, and hence they are valuable for probing distant galaxies. SLSNe have been observed spectroscopically to redshifts up to z $\approx$ 2 \citep{Smith:2018aa}, and photometrically up to z $\approx$ 4 \citep{Cooke:2012aa}.

In spite of the increase in their discovery rate, the SLSNe are rare events. Only $\sim$1 SLSN is observed in a volumetric sample per few thousand SNe \citep{Quimby:2013aa}. This sparsity of sample has led to several unanswered questions regarding the observed diversity in the photometric and spectroscopic properties of the SLSNe and the underlying mechanisms that power these luminous events. The majority of the observed SLSNe are hydrogen-poor in their observed spectra, with strong O II absorption lines early on \citep{Quimby:2011tc,Inserra:2013aa}. These SLSNe are referred to as type-I (SLSN-I) analogous to traditional type I SNe (SN-I). About 150 SLSNe-I have been confirmed spectroscopically \citep{Gomez:2020aa}. The other group of SLSN shows strong, narrow hydrogen emission lines in their spectra, analogous to traditional type IIn SNe (SN-IIn). Despite the smaller sample size, there are well-studied examples of SLSN-II such as SN 2006gy \citep{Ofek:2007aa,Smith:2007aa}, SN 2006tf \citep{Smith:2008vo}, SN 2008fz \citep{Drake:2010aa}, and SN 2008am \citep{Chatzopoulos:2011aa}.

Among the competing theories to explain the immense amount of power required to explain the luminosities of SLSNe, two are more widely accepted given their success in explaining both the luminosities and the timescales of the observed light curves \citep{Nicholl:2021aa}. The first of these is a highly magnetized, rapidly rotating (millisecond) pulsar \citep[magnetar][]{Kasen:2010aa,Woosley:2010aa}. Observed galactic neutron stars have periods larger than a millisecond, however, conservation of angular momentum points to a possibility that they might have been born with a much lower period \citep{Duncan:1992aa}. The other competing theory points to an interaction between slow-moving circumstellar material (CSM) and SN ejecta \citep{Chevalier:1994aa,Smith:2007ab,Chatzopoulos:2012ab,Chevalier:2011aa,Ginzburg:2012aa}. 

The narrow hydrogen emission lines observed in SLSNe-II indicate strong SN ejecta-circumstellar material interaction (hereafter, CSI), as observed in the classical SN-IIn \citep{Schlegel:1990aa}. Spectroscopic models of interacting SNe have been studied mostly in the context of regular--luminosity, H--rich Type IIn events involving collisions between red supergiant (RSG) progenitor SN ejecta and a dense H-rich wind \citep{Dessart:2015wn,Dessart:2016wx,Dessart:2022aa,Dessart:2023aa,Dessart:2023ab,Dessart:2024aa}. These spectroscopic models, which are mainly performed with the CMFGEN code \citep{Dessart:2010aa} have significantly aided the effort to elucidate the properties of the CSM and the progenitor stars, but their application has been limited to regular--luminosity events and H--rich environments. Synthetic spectra for SLSNe interacting with H--deficient CSM of different structural properties (such as density profile and distance from the progenitor stars) are still necessary to further determine the prevalence of CSM interaction for SLSNe and other luminous transients that do not show signs of H in their spectra.

Even though it is difficult to clearly distinguish magnetar model from the CSI as the mechanism that powers the SLSNe \citep{Metzger:2014aa}, \citet{Benetti:2014aa} make an argument that the two spectroscopic classes with and without the hydrogen lines, SLSN-I and SLSN-II could be explained by the same underlying mechanism. In the context of CSS121015, a SLSN-Ic, they found that the observed properties are more consistent with a CSI model over a magnetar model. They further propose that the diversity in the observed properties of different subtypes of SLSNe could be explained in terms of the differences in the structure and hydrogen content of the CSM. 

In this study, we explore the effects on the light curves and spectra of SLSNe for a range of progenitor models combined with a range of circumstellar properties. In section \ref{sec:methods}, we explain the numerical methods employed to evolve the progenitor stars and the SN ejecta. In section \ref{sec:results}, we present the results of our simulations, namely, the evolution of the bolometric luminosity and the spectra for our models. We present a  statistical analysis partially driven by supervised machine learning (ML) techniques performed on the data to find the correlation between the model parameters and the strength of the Balmer series lines. We also present a comparison of our model spectra with the observed SN spectra. Finally, we discuss the conclusions of our study. 

\section{Numerical Methods} \label{sec:methods}

We explored the evolution of the progenitors and the properties of their subsequent SN explosion through the following numerical setup. The evolution of the progenitors was carried out using stellar evolution code Modules for Experiments in Stellar Astrophysics \citep[MESA,][]{Paxton:2011aa,Paxton:2013aa,Paxton:2015aa,Paxton:2018aa,Paxton:2019aa}, followed by the radiation hydrodynamic code {\tt STELLA} \citep{Blinnikov:1993aa,Blinnikov:1998aa,Blinnikov:2004aa,Blinnikov:2006aa} to track the time evolution of the SN ejecta. In the final step, the spectroscopic properties were extracted using a Monte Carlo radiative transfer code {\tt SuperLite}\footnote{Publicly available at Github (\url{https://gururajw.github.io/superlite_docs/}) and Zenodo \citep{Wagle:2023ze} for download. The code documentation is provided at \url{https://gururajw.github.io/superlite_docs/}.} \citep{Wagle:2023aa} by mapping the hydrodynamic profiles of the resulting SN from {\tt STELLA}. We discuss each of these steps  in detail in the following sections. 

\begin{deluxetable*}{|c|ccc|cccccccc}
\tablecaption{Properties of the Progenitors and the CSM. \label{tab:prop}}
\tabletypesize{\footnotesize}
\tablewidth{0pt}
\tablehead{
\colhead{} & \colhead{} & \colhead{} & \colhead{} & \colhead{Model} & \colhead{Explosion} & \colhead{$\rm ^{56}Ni$ mass} & \colhead{$\rm M_{CSM}$} & \colhead{$\rm \dot{M}_{CSM}$} & \colhead{$\rm v_{CSM}$} & \colhead{$\rm t_{CSM}$} & \colhead{CSM} \\[-.2cm]
\colhead{} & \colhead{} & \colhead{} & \colhead{} & \colhead{} & \colhead{Energy} & \colhead{} & \colhead{} & \colhead{} & \colhead{} & \colhead{} \\ [-.2cm]
\colhead{} & \colhead{} & \colhead{} & \colhead{} & \colhead{} & \colhead{[$10^{51}$ erg]} & \colhead{[\msun ]} & \colhead{[\msun ]} & \colhead{[\msun /yr]} & \colhead{[$\rm km s^{-1}$]} & \colhead{[yr]} & \colhead{composition}
}
%\decimalcolnumbers
\startdata
\multirow{15}{*}{\rotatebox[origin=c]{90}{A Series - Red Super Giant}} & \multirow{5}{*}{\rotatebox[origin=c]{90}{Radius}} & \multirow{5}{*}{\rotatebox[origin=c]{90}{at explosion}} & \multirow{5}{*}{\rotatebox[origin=c]{90}{603 \rsun}} & A1 & 0.78 & 0.042 & 0.0 & N/A & N/A & N/A & N/A \\
& & & & A2 & 1.0 & 0.042 & 0.0 & N/A & N/A & N/A & N/A \\
& & & & A3 & 1.2 & 0.042 & 0.0 & N/A & N/A & N/A & N/A \\
& & & & A4 & 0.78 & 0.042 & 0.2 & 0.025 & 200 & 8 & H-rich \\
& & & & A5 & 0.78 & 0.042 & 0.5 & 0.0625 & 200 & 8 & H-rich \\
\cline{2-4}
& \multirow{5}{*}{\rotatebox[origin=c]{90}{Mass}} & \multirow{5}{*}{\rotatebox[origin=c]{90}{at explosion}} & \multirow{5}{*}{\rotatebox[origin=c]{90}{17.8 \msun}} & A6 & 0.78 & 0.042 & 1.0 & 0.125 & 200 & 8 & H-rich \\
& & & & A7 & 0.78 & 0.042 & 1.2 & 0.25 & 200 & 8 & H-rich \\
& & & & A8 & 1.0 & 0.042 & 0.2 & 0.025 & 200 & 8 & H-rich \\
& & & & A9 & 1.0 & 0.042 & 0.5 & 0.0625 & 200 & 8 & H-rich \\
& & & & A10 & 1.0 & 0.042 & 1.0 & 0.125 & 200 & 8 & H-rich \\
\cline{2-4}
& \multirow{5}{*}{\rotatebox[origin=c]{90}{Mass}} & \multirow{5}{*}{\rotatebox[origin=c]{90}{at ZAMS}} & \multirow{5}{*}{\rotatebox[origin=c]{90}{19 \msun}} & A11 & 1.0 & 0.042 & 1.2 & 0.25 & 200 & 8 & H-rich \\
& & & & A12 & 1.2 & 0.042 & 0.2 & 0.025 & 200 & 8 & H-rich \\
& & & & A13 & 1.2 & 0.042 & 0.5 & 0.0625 & 200 & 8 & H-rich \\
& & & & A14 & 1.2 & 0.042 & 1.0 & 0.125 & 200 & 8 & H-rich \\
& & & & A15 & 1.2 & 0.042 & 1.2 & 0.25 & 200 & 8 & H-rich \\
\tableline
\multirow{15}{*}{\rotatebox[origin=c]{90}{B Series - Yellow Super Giant}} & \multirow{5}{*}{\rotatebox[origin=c]{90}{Radius}} & \multirow{5}{*}{\rotatebox[origin=c]{90}{at explosion}} & \multirow{5}{*}{\rotatebox[origin=c]{90}{525 \rsun}} & B1 & 0.28 & 0.009 & 0.0 & N/A & N/A & N/A & N/A \\
& & & & B2 & 0.5 & 0.009 & 0.0 & N/A & N/A & N/A & N/A \\
& & & & B3 & 1.0 & 0.009 & 0.0 & N/A & N/A & N/A & N/A \\
& & & & B4 & 0.28 & 0.009 & 0.2 & 0.025 & 200 & 8 & H-rich \\
& & & & B5 & 0.28 & 0.009 & 0.5 & 0.0625 & 200 & 8 & H-rich \\
\cline{2-4}
& \multirow{5}{*}{\rotatebox[origin=c]{90}{Mass}} & \multirow{5}{*}{\rotatebox[origin=c]{90}{at explosion}} & \multirow{5}{*}{\rotatebox[origin=c]{90}{11.6 \msun}} & B6 & 0.28 & 0.009 & 1.0 & 0.125 & 200 & 8 & H-rich \\
& & & & B7 & 0.28 & 0.009 & 1.2 & 0.25 & 200 & 8 & H-rich \\
& & & & B8 & 0.5 & 0.009 & 0.2 & 0.025 & 200 & 8 & H-rich \\
& & & & B9 & 0.5 & 0.009 & 0.5 & 0.0625 & 200 & 8 & H-rich \\
& & & & B10 & 0.5 & 0.009 & 1.0 & 0.125 & 200 & 8 & H-rich \\
\cline{2-4}
& \multirow{5}{*}{\rotatebox[origin=c]{90}{Mass}} & \multirow{5}{*}{\rotatebox[origin=c]{90}{at ZAMS}} & \multirow{5}{*}{\rotatebox[origin=c]{90}{11.8 \msun}} & B11 & 0.5 & 0.009 & 1.2 & 0.25 & 200 & 8 & H-rich \\
& & & & B12 & 1.0 & 0.009 & 0.2 & 0.025 & 200 & 8 & H-rich \\
& & & & B13 & 1.0 & 0.009 & 0.5 & 0.0625 & 200 & 8 & H-rich \\
& & & & B14 & 1.0 & 0.009 & 1.0 & 0.125 & 200 & 8 & H-rich \\
& & & & B15 & 1.0 & 0.009 & 1.2 & 0.25 & 200 & 8 & H-rich \\
\tableline
\multirow{15}{*}{\rotatebox[origin=c]{90}{C Series - Blue Super Giant}} & \multirow{5}{*}{\rotatebox[origin=c]{90}{Radius}} & \multirow{5}{*}{\rotatebox[origin=c]{90}{at explosion}} & \multirow{5}{*}{\rotatebox[origin=c]{90}{7.24 \rsun}} & C1 & 0.80 & 0.011 & 0.0 & N/A & N/A & N/A & N/A \\
& & & & C2 & 0.95 & 0.011 & 0.0 & N/A & N/A & N/A & N/A \\
& & & & C3 & 1.50 & 0.011 & 0.0 & N/A & N/A & N/A & N/A \\
& & & & C4 & 0.80 & 0.011 & 0.2 & 0.2 & 1000 & 1 & He-rich \\
& & & & C5 & 0.80 & 0.011 & 0.4 & 0.4 & 1000 & 1 & He-rich \\
\cline{2-4}
& \multirow{5}{*}{\rotatebox[origin=c]{90}{Mass}} & \multirow{5}{*}{\rotatebox[origin=c]{90}{at explosion}} & \multirow{5}{*}{\rotatebox[origin=c]{90}{3.4 \msun}} & C6 & 0.80 & 0.011 & 1.0 & 1.0 & 1000 & 1 & He-rich \\
& & & & C7 & 0.80 & 0.011 & 2.0 & 2.0 & 1000 & 1 & He-rich \\
& & & & C8 & 0.95 & 0.011 & 0.2 & 0.2 & 1000 & 1 & He-rich \\
& & & & C9 & 0.95 & 0.011 & 0.4 & 0.4 & 1000 & 1 & He-rich \\
& & & & C10 & 0.95 & 0.011 & 1.0 & 1.0 & 1000 & 1 & He-rich \\
\cline{2-4}
& \multirow{5}{*}{\rotatebox[origin=c]{90}{Mass}} & \multirow{5}{*}{\rotatebox[origin=c]{90}{at ZAMS}} & \multirow{5}{*}{\rotatebox[origin=c]{90}{11 \msun}} & C11 & 0.95 & 0.011 & 2.0 & 2.0 & 1000 & 1 & He-rich \\
& & & & C12 & 1.50 & 0.011 & 0.2 & 0.2 & 1000 & 1 & He-rich \\
& & & & C13 & 1.50 & 0.011 & 0.4 & 0.4 & 1000 & 1 & He-rich \\
& & & & C14 & 1.50 & 0.011 & 1.0 & 1.0 & 1000 & 1 & He-rich \\
& & & & C15 & 1.50 & 0.011 & 2.0 & 2.0 & 1000 & 1 & He-rich
\enddata
\tablecomments{The H-rich CSM composition contains X(H) $\approx$ 0.64 and X(He) $\approx$ 0.34, while the He-rich CSM composition contains X(H) $\approx$ 0 and X(He) $\approx$ 0.84. The composition of the CSM is set to be the same as the surface composition of the progenitor star based on the results of the corresponding MESA simulation.} 
\end{deluxetable*}

\subsection{Progenitor evolution}

We calculated the evolution of three isolated massive progenitor stars using \texttt{MESA}. The progenitor stars were evolved with initial properties chosen to produce a red supergiant (RSG; with the supernova mass, $\rm M_{SN}$ = 17.8 \msun \ and the radius $\rm R_{SN}$ = 603 \rsun), a yellow supergiant (YSG; $\rm M_{SN}$ = 11.6 \msun, $\rm R_{SN}$ = 525 \rsun) and a blue supergiant (BSG; $\rm M_{SN}$ = 3.4 \msun, $\rm R_{\rm SN}$ = 7.24 \rsun) as a pre-explosion model. This choice was made to investigate the effects of SN ejecta mass and composition, and the CSM composition, on the resulting spectral evolution for a variety of different initial conditions. For all the computed {\tt MESA} models, we adopted solar metallicity \citep{Asplund:2009aa} at Zero Age Main Sequence, no rotation, a Ledoux criterion for convection with the mixing-length parameter $\alpha_{\rm MLT}$=1.5, a 21-isotope nuclear reaction network, and the ``Dutch" mass-loss prescription \citep{Nugis:2000aa,Vink:2001aa,Glebbeek:2009aa} that is recommended for massive stars within our target SN progenitor mass range\footnote{The inlist files are available on Zenodo under an open-source  Creative Commons Attribution license:  \dataset[doi:10.5281/zenodo.13890925]{https://doi.org/10.5281/zenodo.13890925}.}.

For each of the three progenitors, we used the same method described in \citet{Paxton:2018aa} to produce 1D, spherically symmetric SN explosions corresponding to three different SN explosion energy choices for each model (low, medium, and high SN explosion energy; see Table \ref{tab:prop}). The composition of the CSM is set to be the same as the surface composition of the progenitor star that resulted from the MESA simulation, as the CSM consists of the steady-state winds produced in the latest stages of the stellar evolution. In each case, we evolved the SN shock wave until the point before the shock break-out from the stellar surface. Subsequently, for the majority of the resulting pre-explosion models, we appended wind-like CSM structures following the $r^{-2}$ density law corresponding to different mass-loss rates ($\dot{\rm M}_{\rm CSM}$) and wind velocities ($\rm v_{CSM}$) which, in turn, translate to different mass--loss timescales ($\rm t_{CSM}$) and total CSM masses ($\rm M_{CSM}$). The CSM properties listed in Table \ref{tab:prop} for each model series were set to represent the typical properties based on the types of progenitors they represent. Model C progenitors are akin to WR-star progenitors with strong winds that have lost their H envelopes. For these compact stars, characteristic wind speeds are typically higher ($\sim$1,000 km/s), and mass loss timescales are typically lower (M/$\dot{\rm M}$ decreases for an increasing $\dot{\rm M}$). The mass-loss takes place at a later stage, therefore the total radius of the CSM created assuming steady-state winds is $\sim 10^{15} \ cm$; shorter than that typical of RSGs and BSGs (which can extend to $\sim \times 10^{16} \ cm$). This resulted in a total of 45 models of different initial progenitor and CSM conditions, with their properties detailed in Table~\ref{tab:prop}. From hereafter, we will refer to the RSG progenitor models as ``Series A", the YSG progenitor models as ``Series B" and the BSG progenitor models as ``Series C".

All 45 explosions were then processed with the two-temperature radiation transfer code {\tt STELLA} \citep{Blinnikov:1998aa,Blinnikov:2006aa}. {\tt STELLA} solves the radiative transfer equations in the intensity momentum approximation in each frequency bin and computes bolometric and broad-band light curves of SN explosions, including for those affected by the interaction with a CSM. {\tt STELLA} is offered within each {\tt MESA} code release, allowing for a smooth transition between the two codes to facilitate the modeling of SN explosions. {\tt STELLA} also yields radiation hydrodynamics profiles for several phases following the explosion, providing the inputs we need to compute time-series of synthetic spectra for each SN model processed.

\subsection{Radiative transfer calculations with SuperLite}

We use the hydrodynamic profiles generated by {\tt STELLA} at several checkpoints through about 30-50 days since the beginning of the {\tt STELLA} simulation for the three model series. The profiles are processed with the time-independent radiation transport code SuperLite. The profiles are truncated at a Rosseland mean optical depth of 100, as computed by {\tt STELLA}. As noted in \citet{Wagle:2023aa}, an optical depth of 30-50 is sufficient for correctly predicting the transitions of lines in the region of interest. We choose a value of 100 to be exhaustive. We choose 6000 wavelength groups with a range of 1 to 30,000 \text{\AA} for the radiative transfer calculations. The hydrodynamic profiles contain the kinetic and radiation temperature profiles, along with the other properties. About a million ($2^{20}$) MC particles are instantiated for the simulation using the radiation temperature provided in the hydrodynamic profile. The spectra are evaluated assuming local thermodynamic equilibrium (LTE) for the line opacities and emissivities. The ionization balance is calculated using the Saha equation and the excitation level populations are calculated using Boltzmann statistics under the LTE assumption, using the electron temperature. We also calculate the spectra for non-LTE (NLTE) conditions by solving the rate equation matrix for hydrogen line transitions as described in \citet{Wagle:2023aa}.

\section{Results} \label{sec:results}

In this section, we present the results of our hydrodynamic and radiative transfer simulations for the three series of models: A, B, and C. The following subsections discuss the photometric and spectroscopic evolution of these models and the significance of the CSM interaction. We then analyze the fluxes of the dominant H Balmer series lines \halpha \ and \hbeta , and their ratio, F(\halpha)/F(\hbeta). In the final subsection, we compare the spectral synthesis results of the {\tt SuperLite} code to the observed spectra.

\subsection{Light curve evolution} \label{subsec:res_lc}

\begin{figure*}[ht!]
\begin{center}
\includegraphics[width=0.92\textwidth]{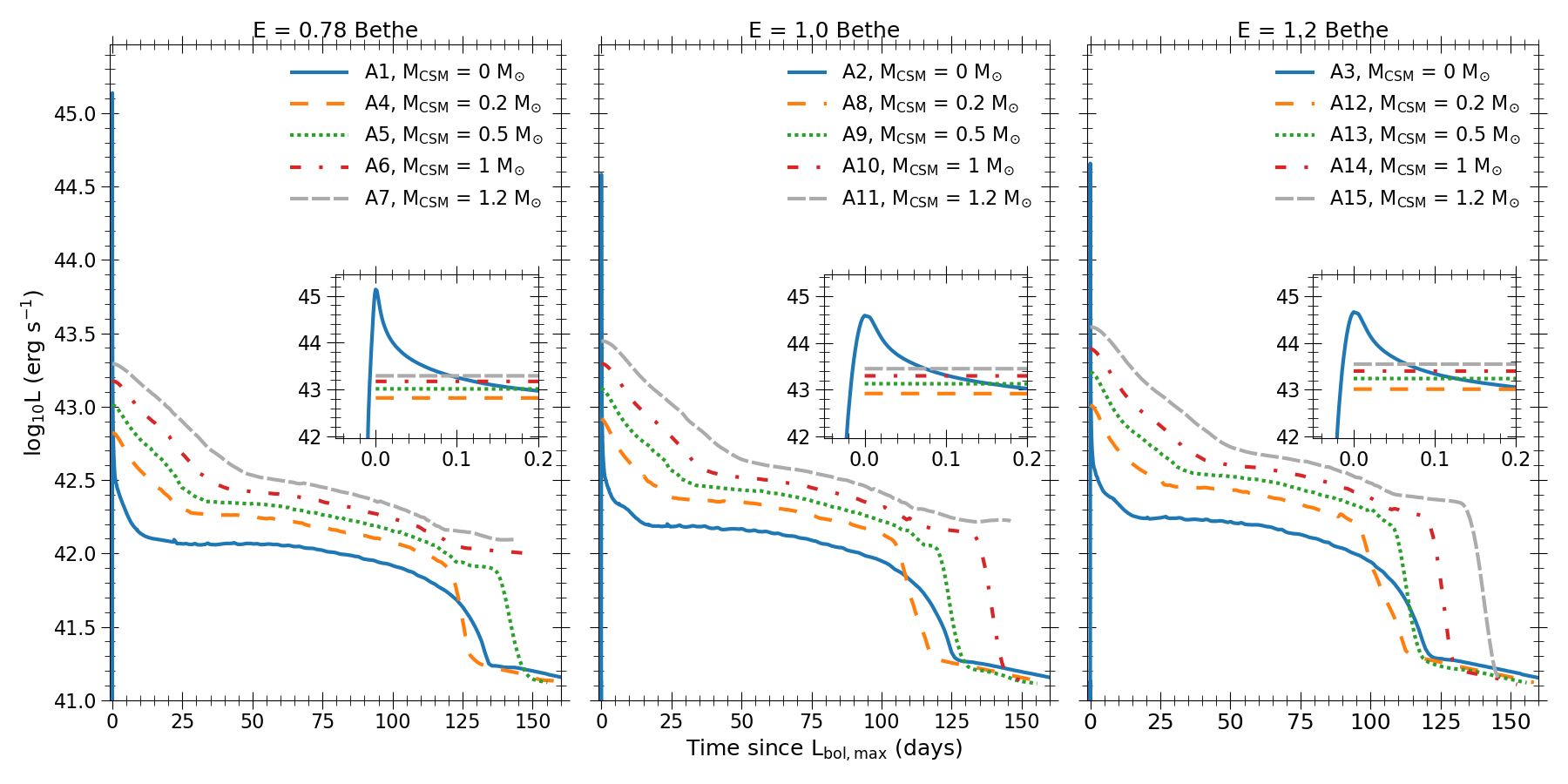}
\caption{The light curves for the A-Series models representing RSG progenitors are shown here for a range of CSM masses and explosion energy values. The properties of the progenitor and the CSM for each of these models are listed in Table \ref{tab:prop}.
\label{fig:LC_A}}
\end{center}
\end{figure*}

\begin{figure*}[ht!]
\begin{center}
\includegraphics[width=0.92\textwidth]{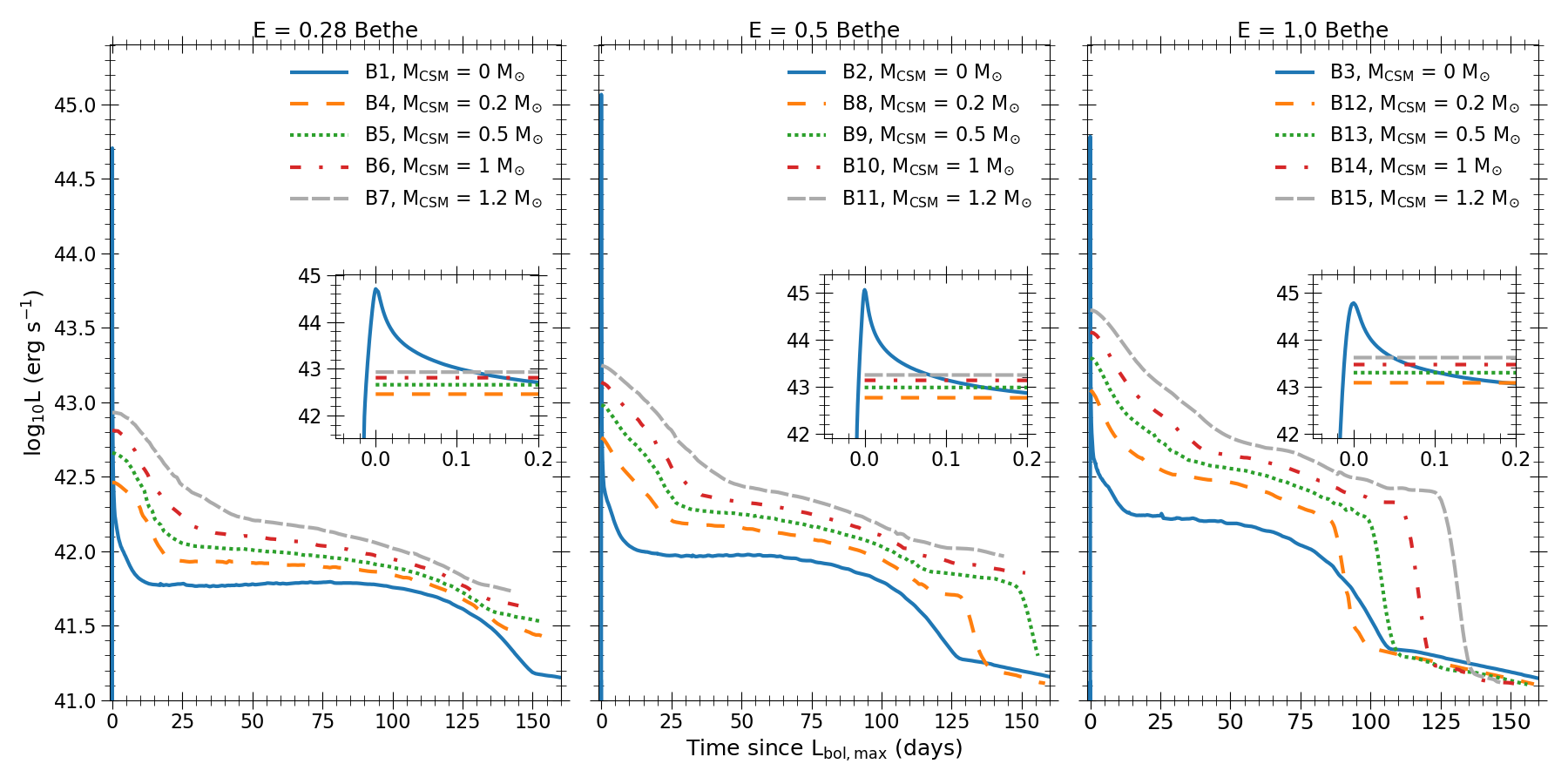}
\caption{The light curves for the B-Series models representing YSG progenitors are shown here for a range of CSM masses and explosion energy values. The insets show light curves near the maximum value. The properties of the progenitor and the CSM for each of these models are listed in Table \ref{tab:prop}.
\label{fig:LC_B}}
\end{center}
\end{figure*}

\begin{figure*}[ht!]
\begin{center}
\includegraphics[width=0.92\textwidth]{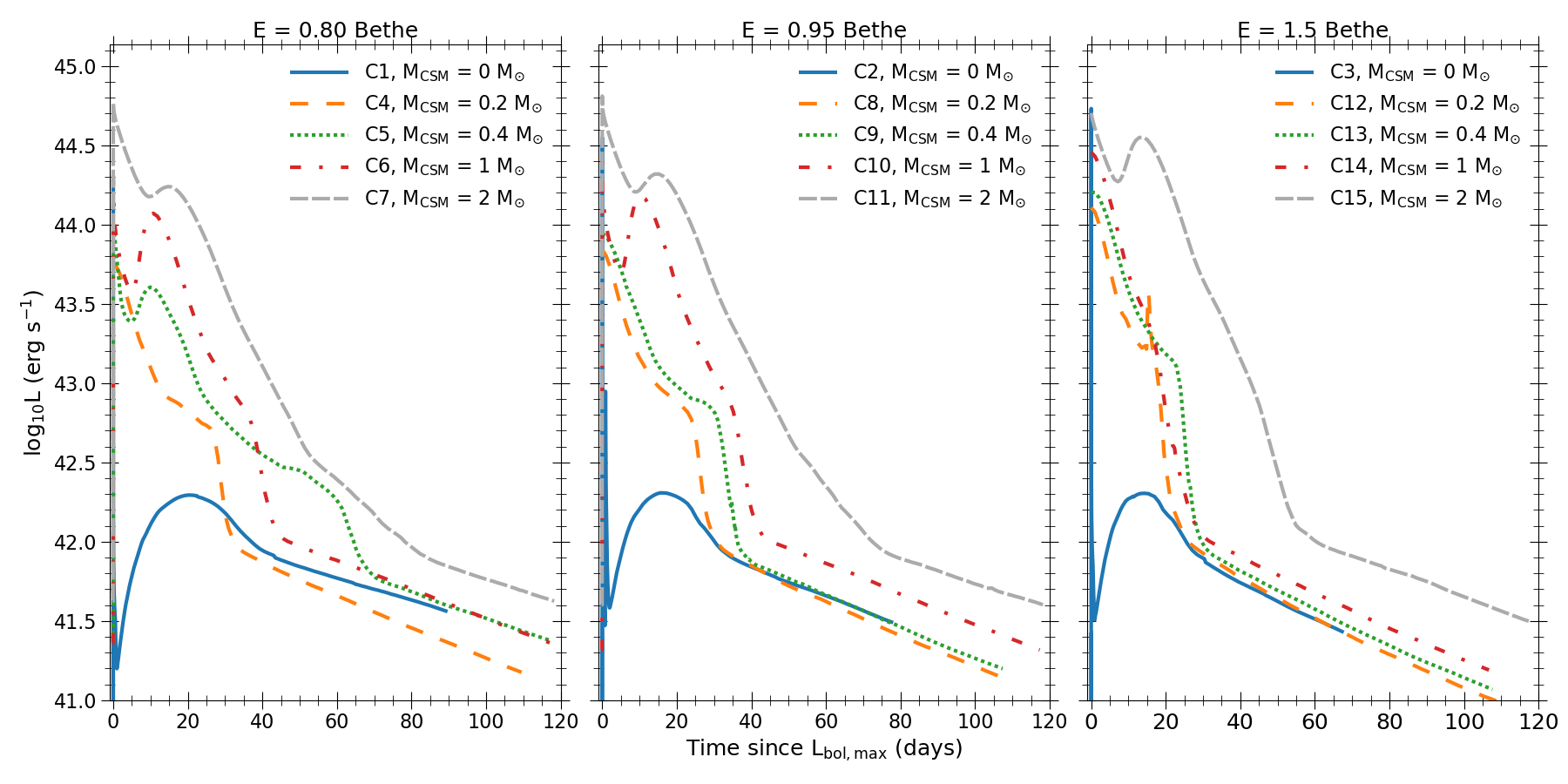}
\caption{The light curves for the C-Series representing BSG progenitors models are shown here for a range of CSM masses and explosion energy values. The insets show light curves near the maximum value. The properties of the progenitor and the CSM for each of these models are listed in Table \ref{tab:prop}.
\label{fig:LC_C}}
\end{center}
\end{figure*}

In Figures \ref{fig:LC_A} to \ref{fig:LC_C}, we present the bolometric light curve evolution of our models for series A, B, and C, respectively. For each series, the models are grouped together with the explosion energy used in the {\tt STELLA} simulation. The figures exhibit the differences in the morphology of the light curves based on the combination of the progenitor and the CSM properties. For the A-series (RSG) and B-series (YSG) models, we see a rise in the luminosity followed by a plateau phase characteristic of SN-IIP. The plateau phase lasts for 80-120 days, representing the recombination of hydrogen in the H-rich envelope, following which the luminosity drops until the radioactive tail is reached at around 100-140 days. The models with no CSM envelope show luminosity declining right after the initial shock breakout reaching a plateau phase. For the models that have a CSM envelope, the luminosity rises slowly after the initial shock-breakout and reaches a maximum in luminosity at a few tens of days. The rise time to the peak, the peak luminosity, and the decline time after the peak to reach the plateau phase increase with increasing CSM mass. As expected, the higher explosion energy produces brighter light curves. Peak luminosities of a few times $\rm 10^{42}-10^{43} \ erg \ s^{-1}$, typical of both standard Type IIn SNe but also more luminous SLSN-II are reached for these models.

For the C-series (BSG) models without CSM envelope, the light curves are morphologically similar to SN-Ib/c, showing a luminosity peak powered by the radioactive decay of $^{56}\rm Ni$ after the initial shock breakout, followed by a steep decline. For the models with a CSM envelope, the energy deposited by the shocked CSM powers the light curve. The luminosity declines for 10-15 days, reaching a second peak around 10-20 days, which is when the radioactive decay of $^{56}\rm Ni$ heats the underlying ejecta. The second peak for the models with larger CSM is brighter than the models with lower or no CSM, relative to their primary peak. These model light curves morphologically resemble those of the classical SN-Ib; however, the peak luminosities of both the primary and secondary peaks reach a few times $\rm 10^{43}-10^{44} \ erg \ s^{-1}$, typical of more luminous SLSN-I.

\subsection{Spectroscopic evolution} \label{subsec:res_spectra}

\begin{figure*}[ht!]
\begin{center}
\gridline{\fig{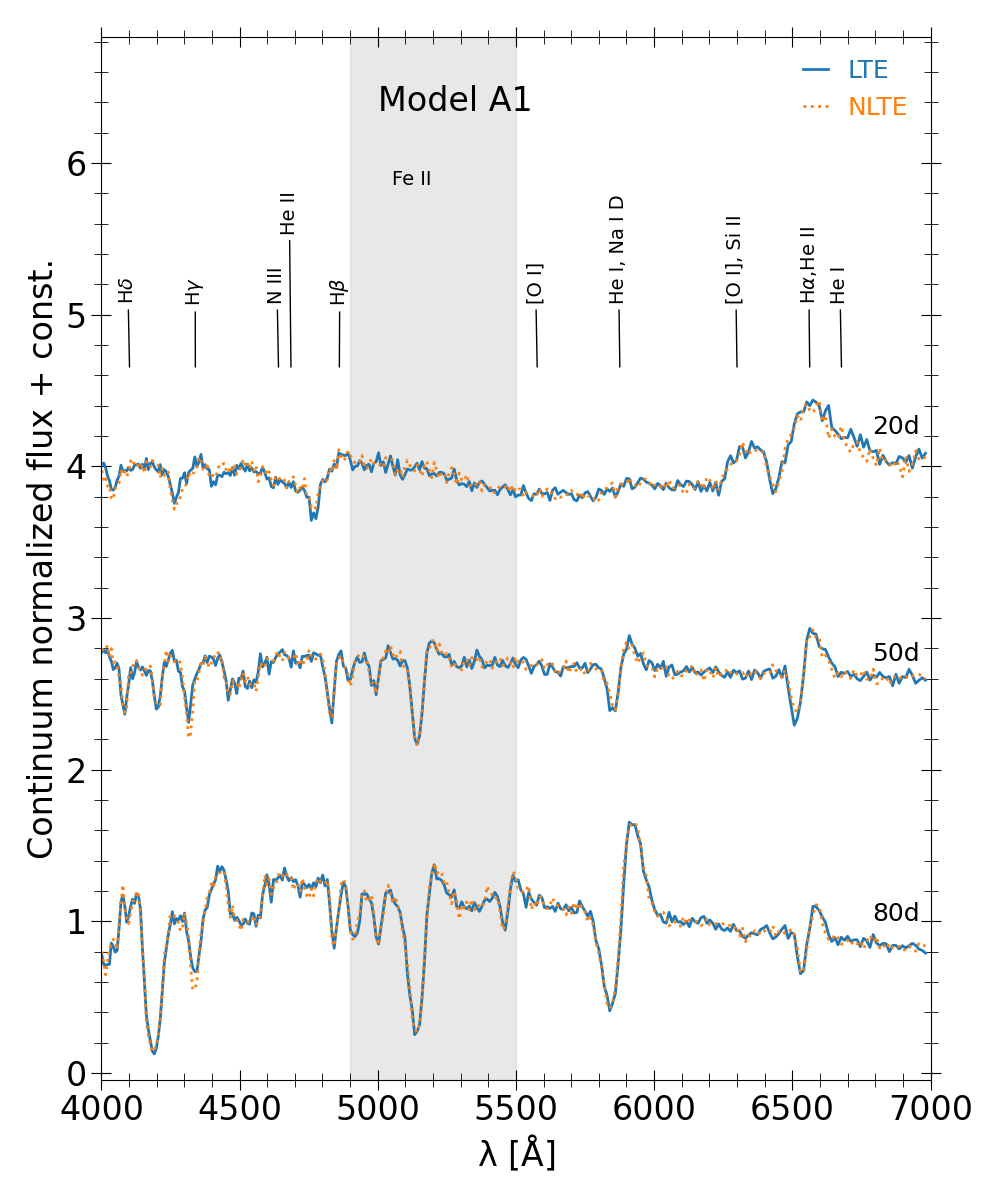}{0.35\textwidth}{(a)}
         \fig{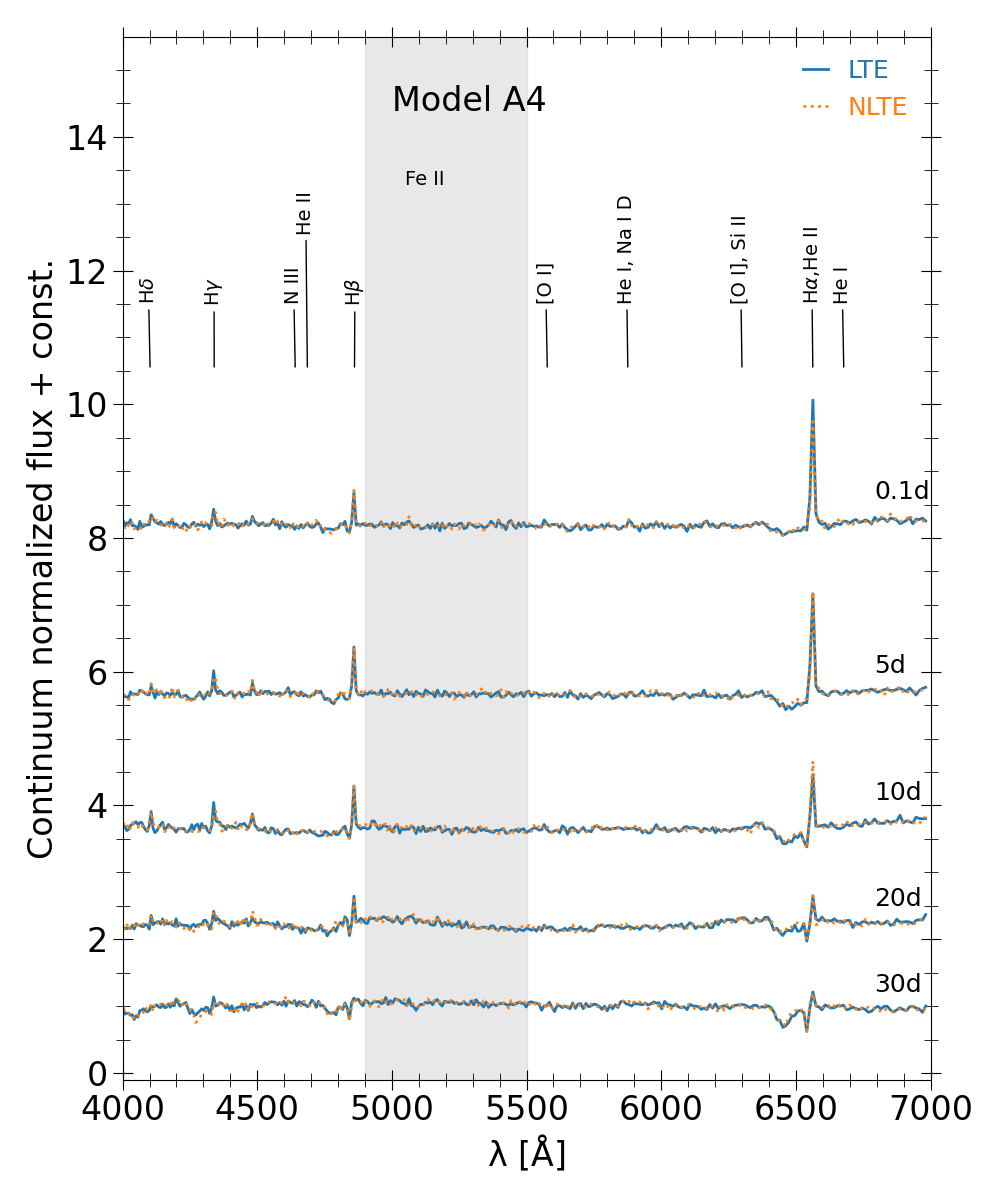}{0.35\textwidth}{(b)}}
\gridline{\fig{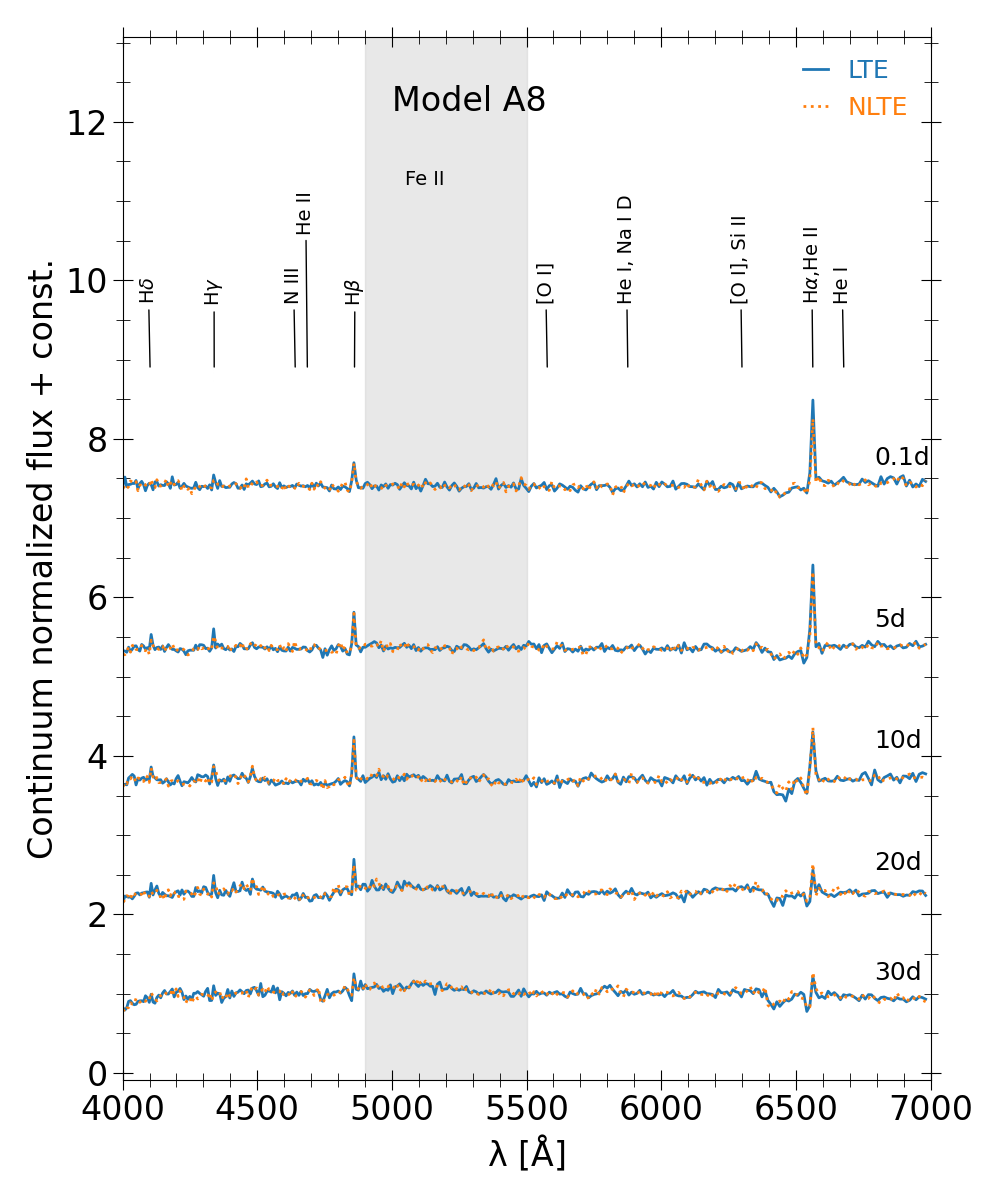}{0.35\textwidth}{(c)}
         \fig{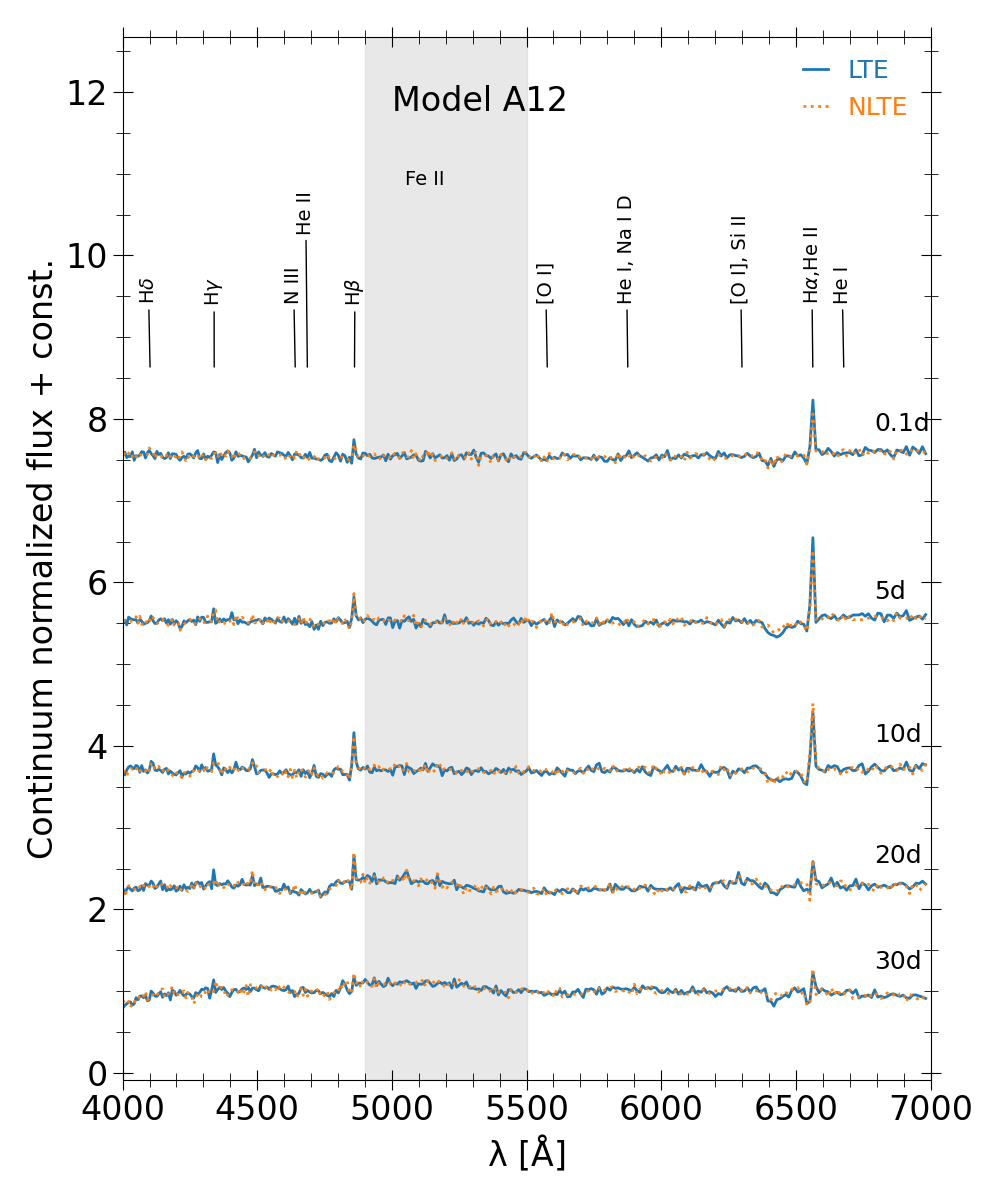}{0.35\textwidth}{(d)}}
\caption{The time evolution of spectra produced by the SuperLite code for some of the A-Series models representing RSG progenitors is shown here. The properties of these models are listed in Tables \ref{tab:prop} \& \ref{tab:A_Series}. Model A1 where the CSM is absent shows typical P-Cygni lines of hydrogen for the phases representing the plateau phase of the light curve, while the other models where the CSM is present show strong emission lines of hydrogen, which diminish in strength as time progresses. \label{fig:spectra_A_Series}}
\end{center}
\end{figure*}

\begin{figure*}[ht!]
\begin{center}
\gridline{\fig{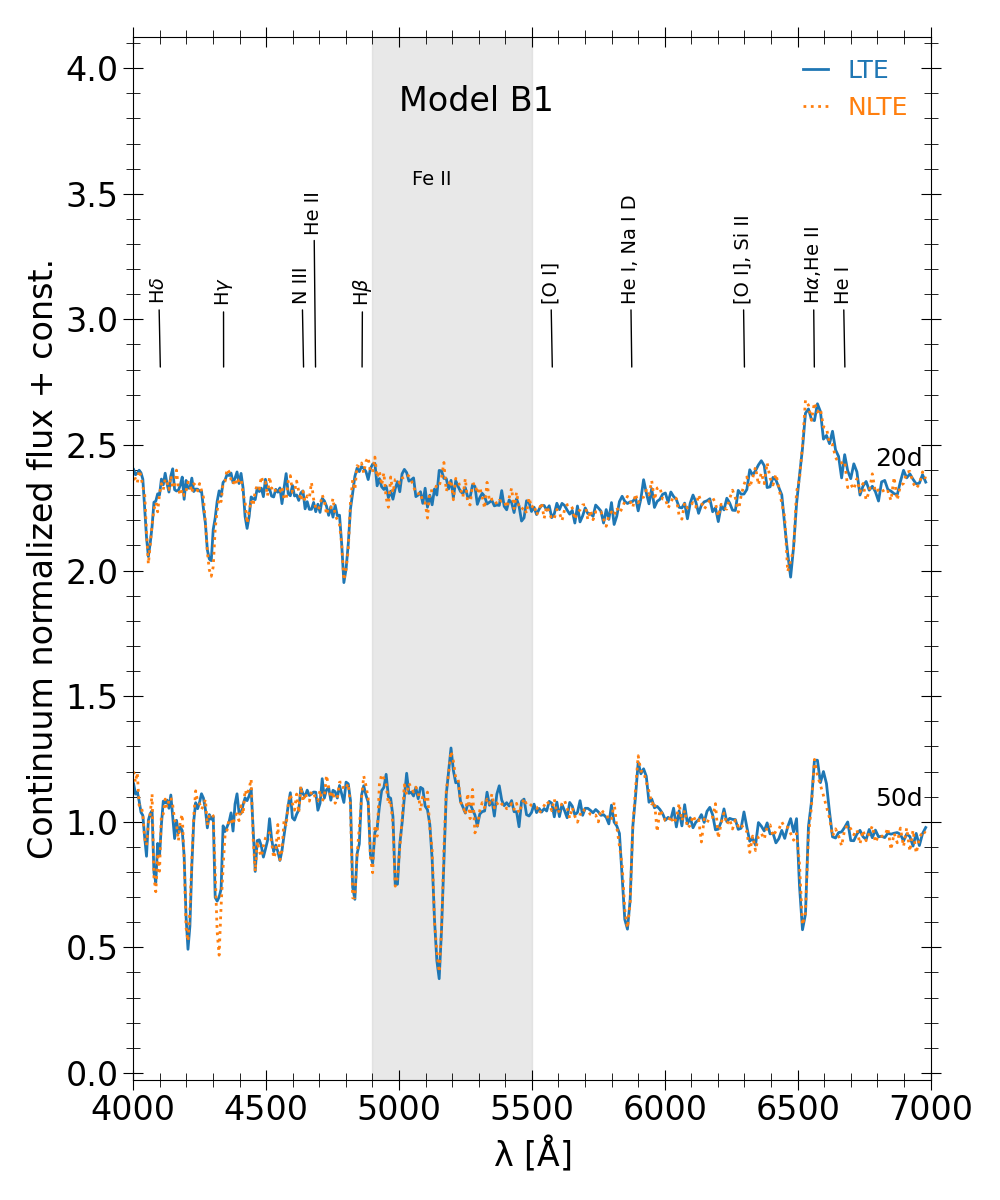}{0.35\textwidth}{(a)}
         \fig{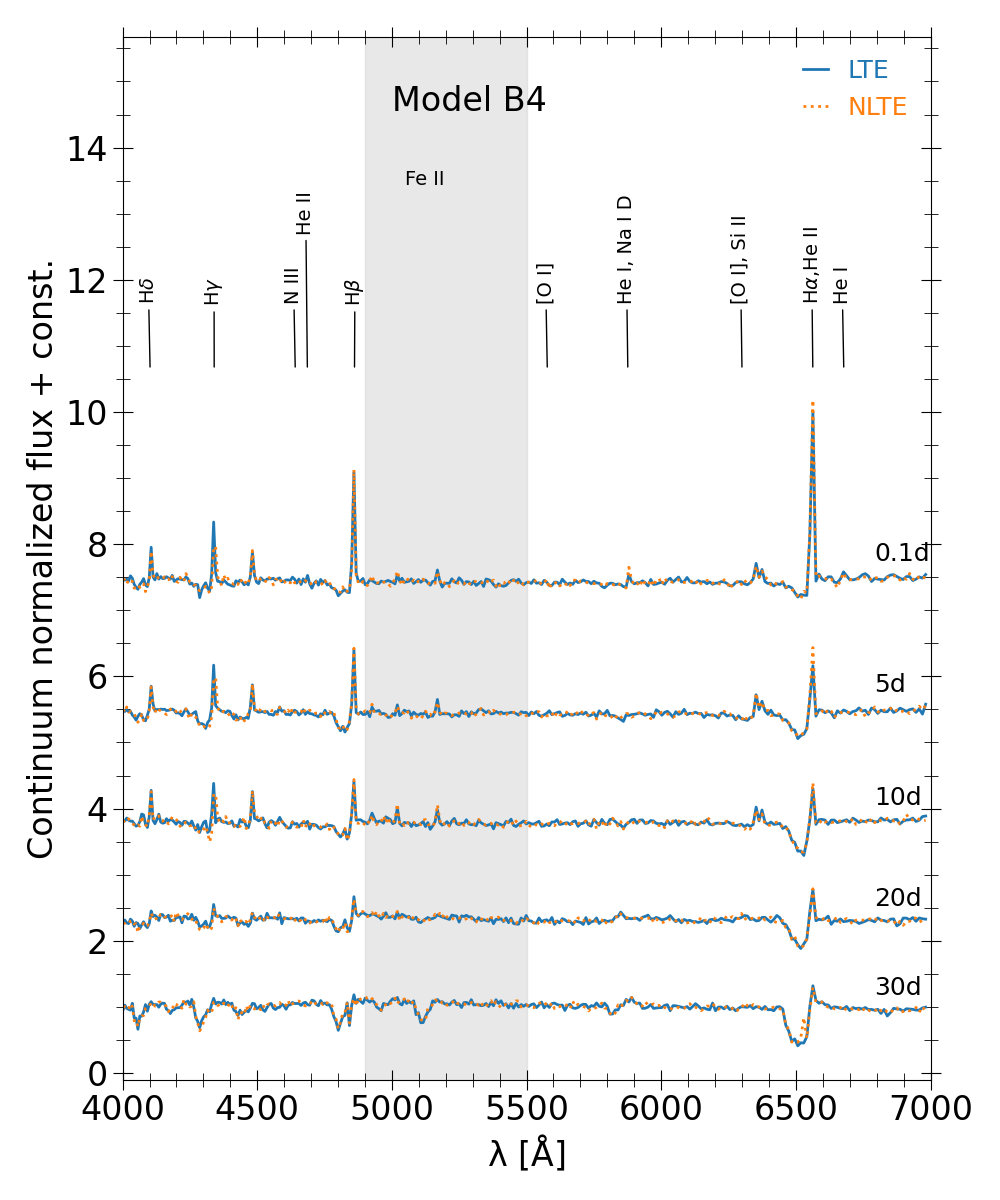}{0.35\textwidth}{(b)}}
\gridline{\fig{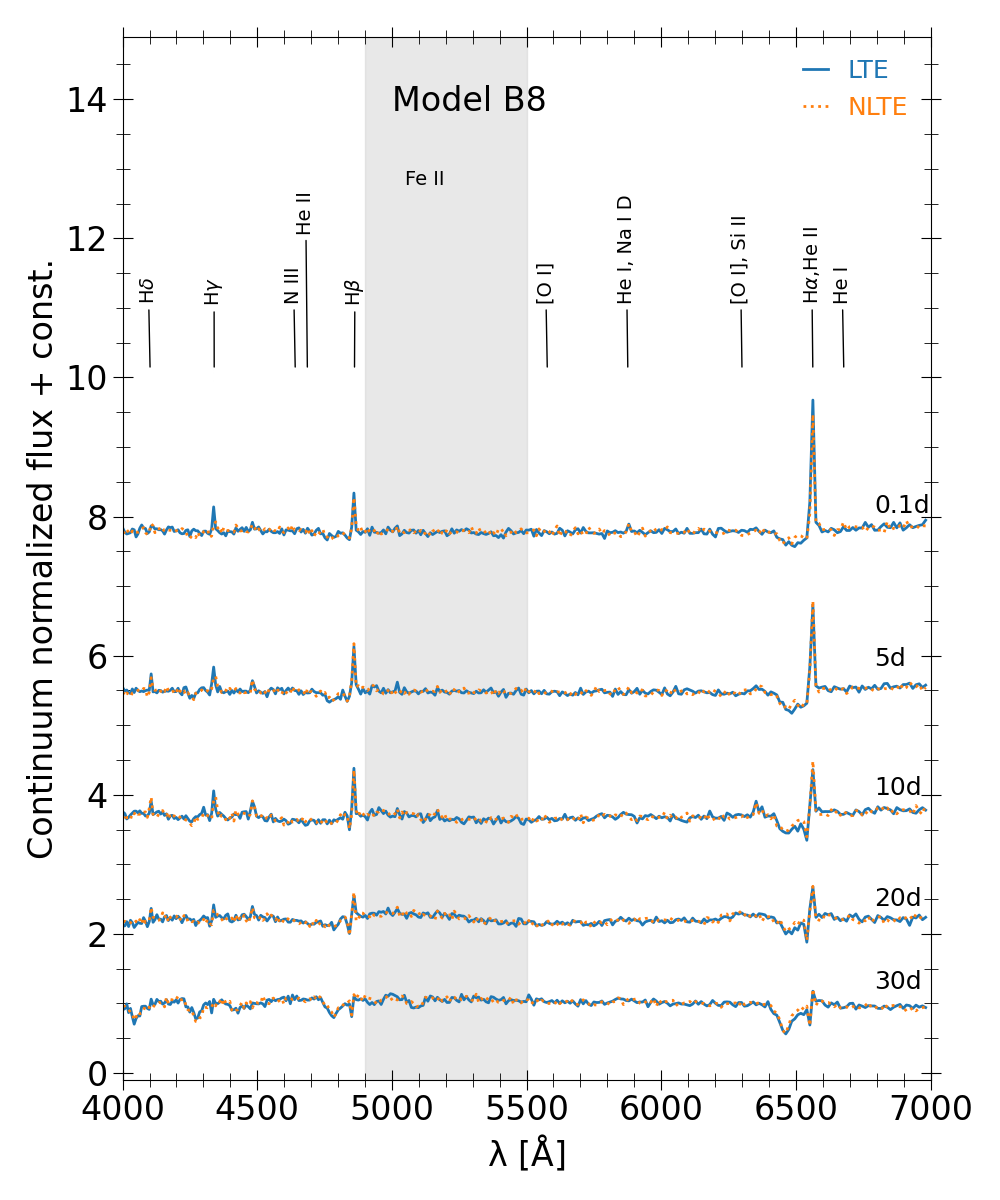}{0.35\textwidth}{(c)}
         \fig{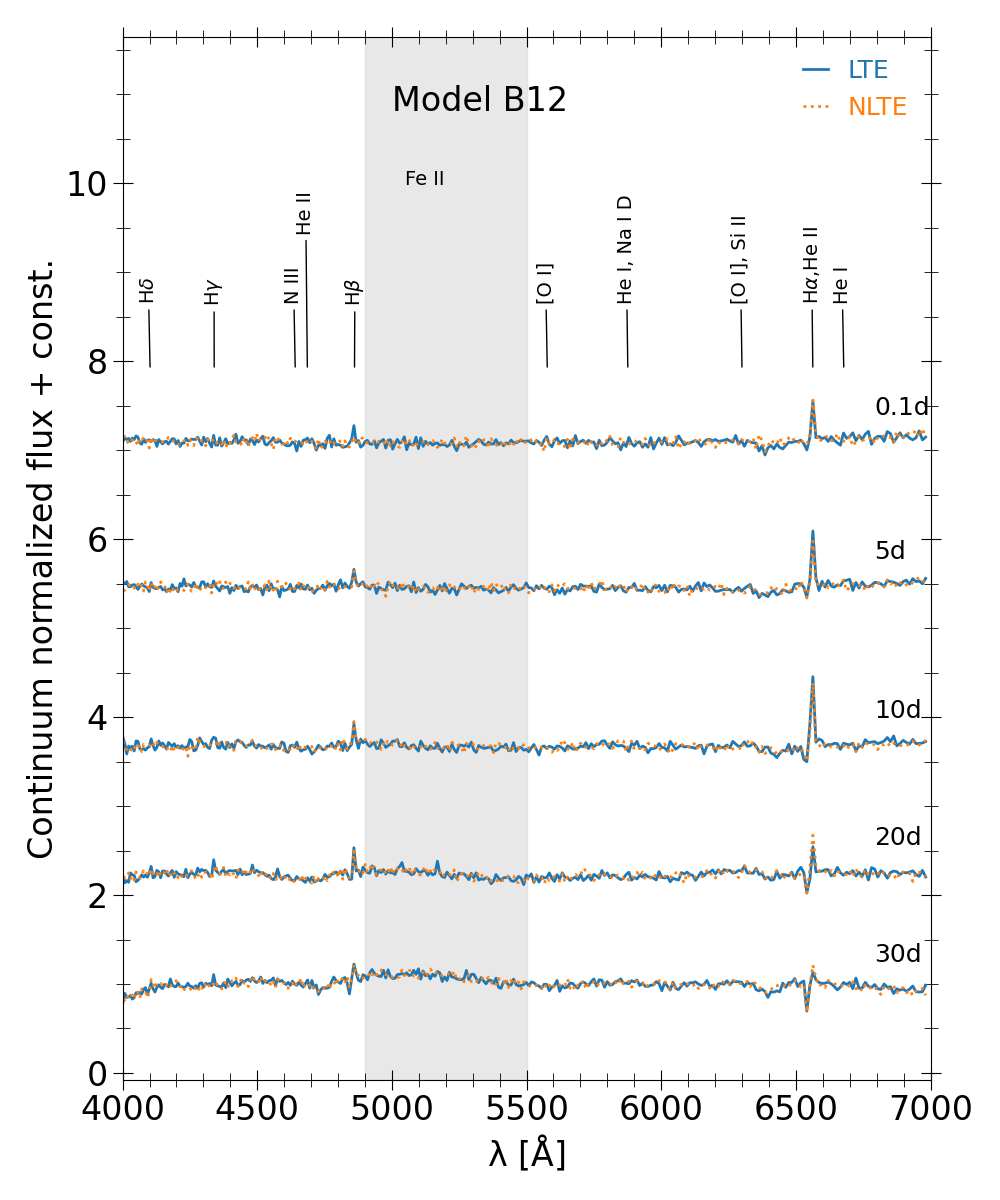}{0.35\textwidth}{(d)}}
\caption{The time evolution of spectra produced by the SuperLite code for some of the B-Series models representing YSG progenitors is shown here. The properties of these models are listed in Tables \ref{tab:prop} \& \ref{tab:B_Series}. Model B1 where the CSM is absent shows typical P-Cygni lines of hydrogen, while the other models where the CSM is present show strong emission lines of hydrogen similar to A series models shown in Fig. \ref{fig:spectra_A_Series} \label{fig:spectra_B_Series}}
\end{center}
\end{figure*}

\begin{figure*}[ht!]
\begin{center}
\gridline{\fig{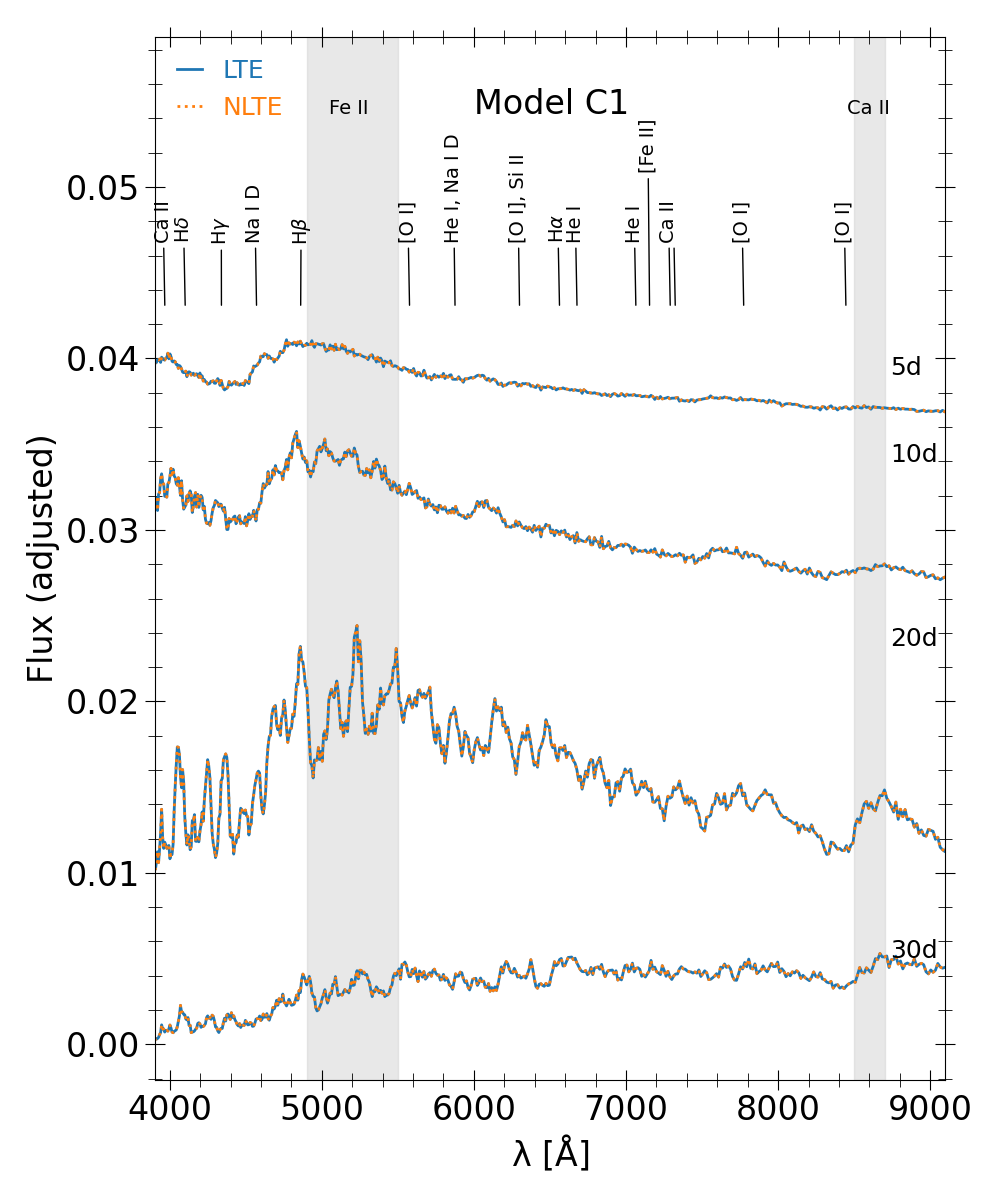}{0.33\textwidth}{(a)}
         \fig{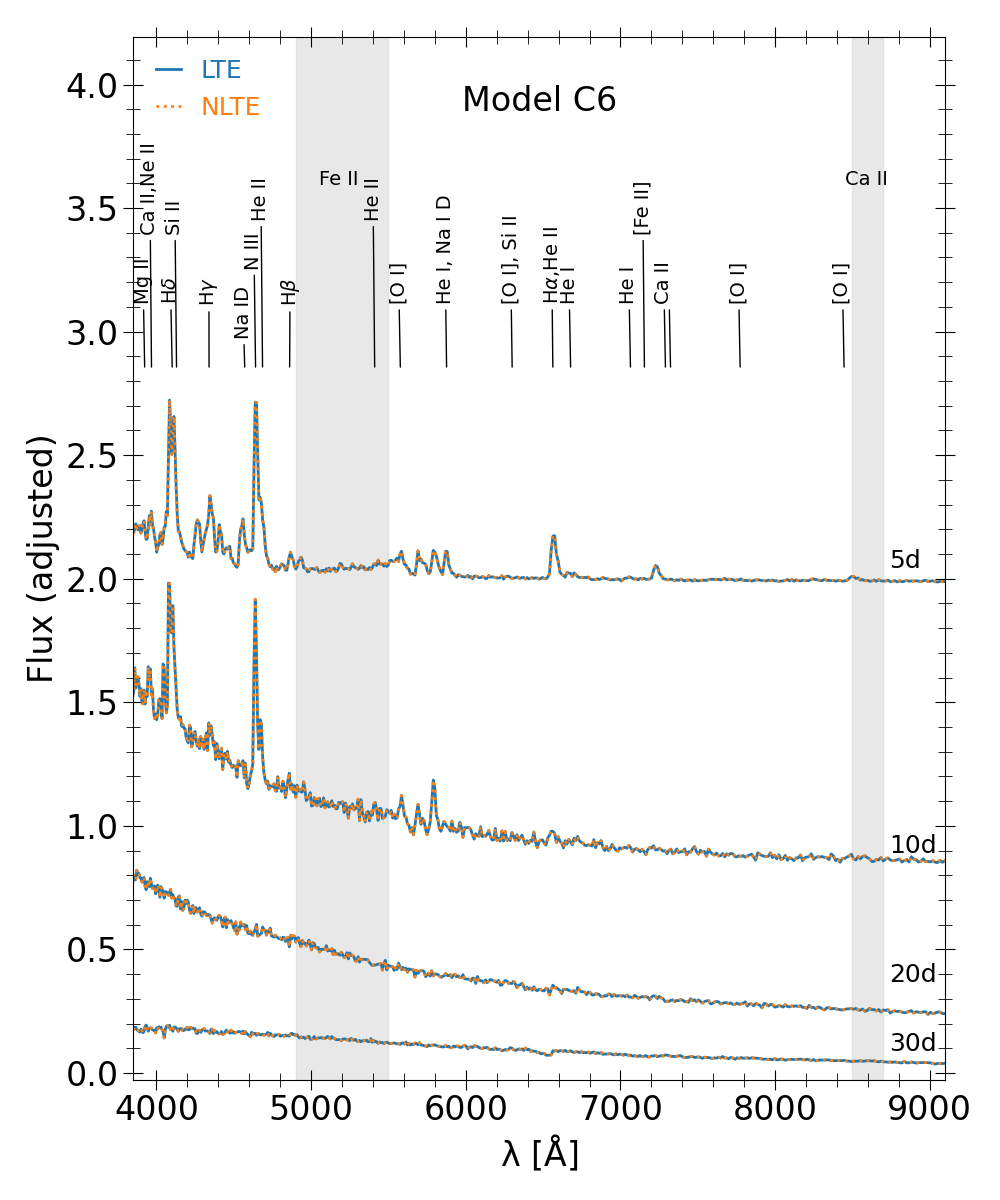}{0.33\textwidth}{(b)}
         \fig{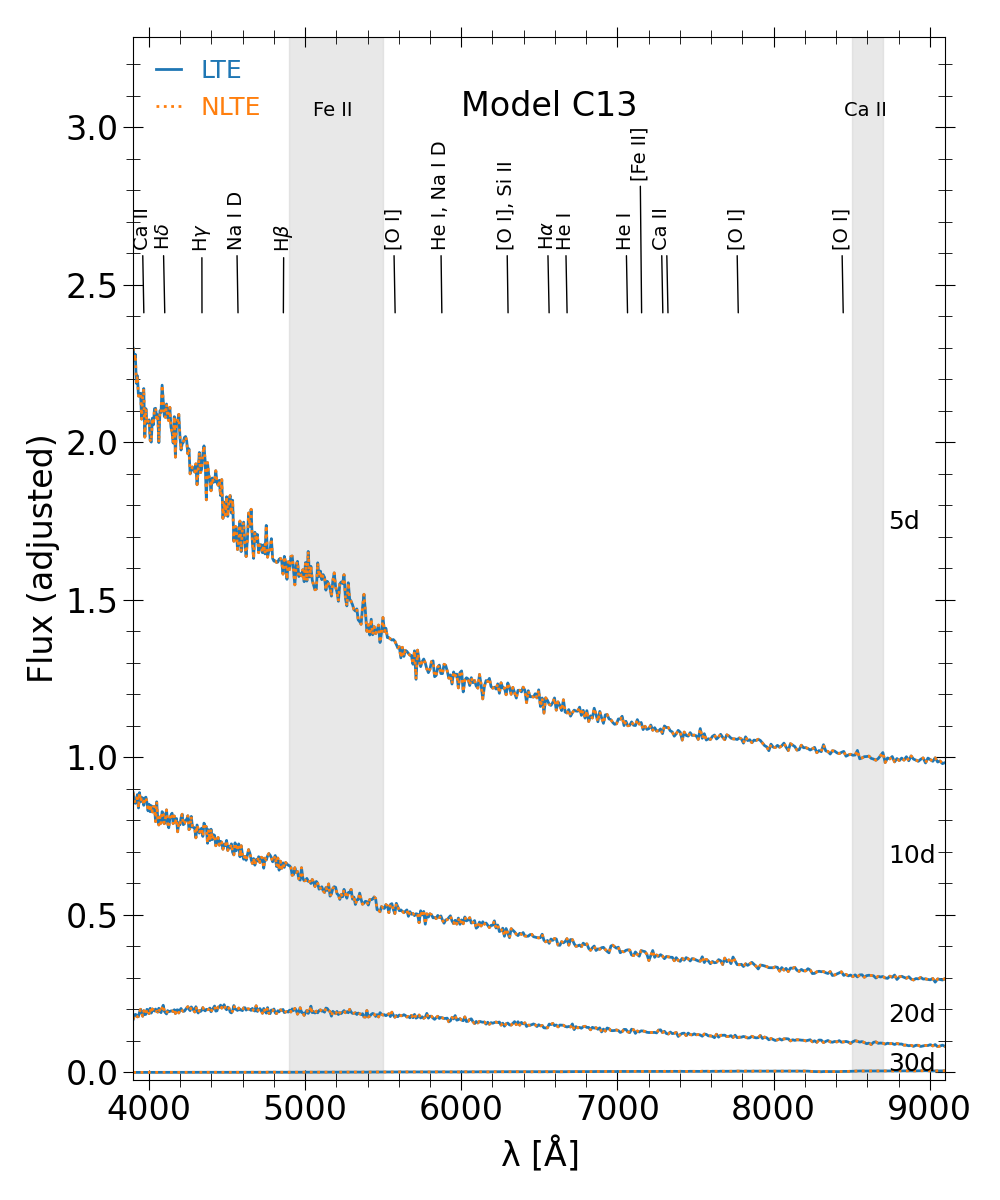}{0.33\textwidth}{(c)}}
\caption{The time evolution of spectra produced by the SuperLite code for some of the C-Series models representing BSG progenitors is shown here. The properties of these models are listed in Tables \ref{tab:prop}. As hydrogen is absent in the ejecta and CSM of the C-series models, the NLTE treatment for hydrogen does not affect the results. The NLTE model spectra are included to emphasize the absence of hydrogen lines in the spectra %\& \ref{tab:C_Series}. 
\label{fig:spectra_C_Series}}
\end{center}
\end{figure*}

The model spectra for the A- through C-series models are displayed in Figures \ref{fig:spectra_A_Series} to \ref{fig:spectra_C_Series}, respectively. The spectra in Figures \ref{fig:spectra_A_Series} \& \ref{fig:spectra_B_Series} are normalized to the continuum using the Specutils package \citep{specutil_2022} in Python. A generalized continuum is fitted in the range of 400-700 nm, excluding the region of strong emission lines. The figures show the evolution of spectra between 5 days to a few tens of days since the peak luminosity. In the models where the CSM envelope is absent, the hydrogen lines are broader as expected for rapidly expanding SN ejecta. As time progresses, the P Cygni profile developing for the lines can be observed in these models. This behavior is of course typical of SN IIP. The He lines also get stronger at this epoch due to the deeper layers of the ejecta being traced by the photosphere. The metal lines for iron and calcium start appearing in the spectra of these models at later times.

The Series A and Series B models with dense H--rich CSM winds exhibit strong, narrow hydrogen lines in their spectra, as is typical for SN IIn. The properties of the Balmer series hydrogen lines \halpha \ and \hbeta \ are listed in Tables \ref{tab:A_Series} and \ref{tab:B_Series} for the A- and B-series models, respectively. The strength of the \halpha \ and the \hbeta \ lines drop as time progresses. At any epoch relative to peak luminosity, the F(\halpha)/F(\hbeta) ratio is higher for models with greater CSM mass compared to those with lower or no CSM. The models show similar trends in the time evolution of the two Balmer series hydrogen lines to the SN models of \citet{Dessart:2016wx} that include the CSI, even though their CSM properties are vastly different (see, for example, the spectra of their model A in figure 5.) The strength of the lines diminishes as time progresses, eventually showing P-Cygni features at later times. 

\begin{figure*}[ht!]
\begin{center}
\gridline{\fig{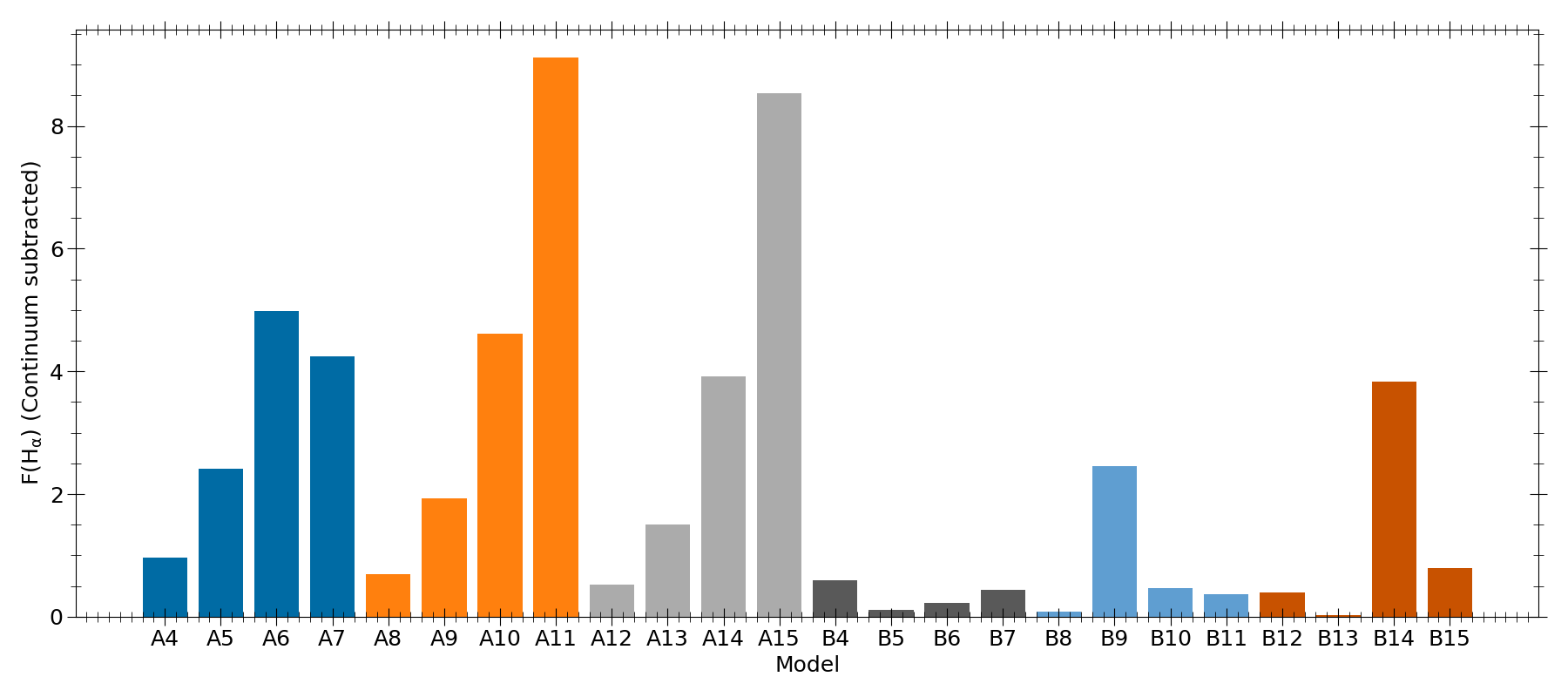}{0.85\textwidth}{(a)}}
\gridline{\fig{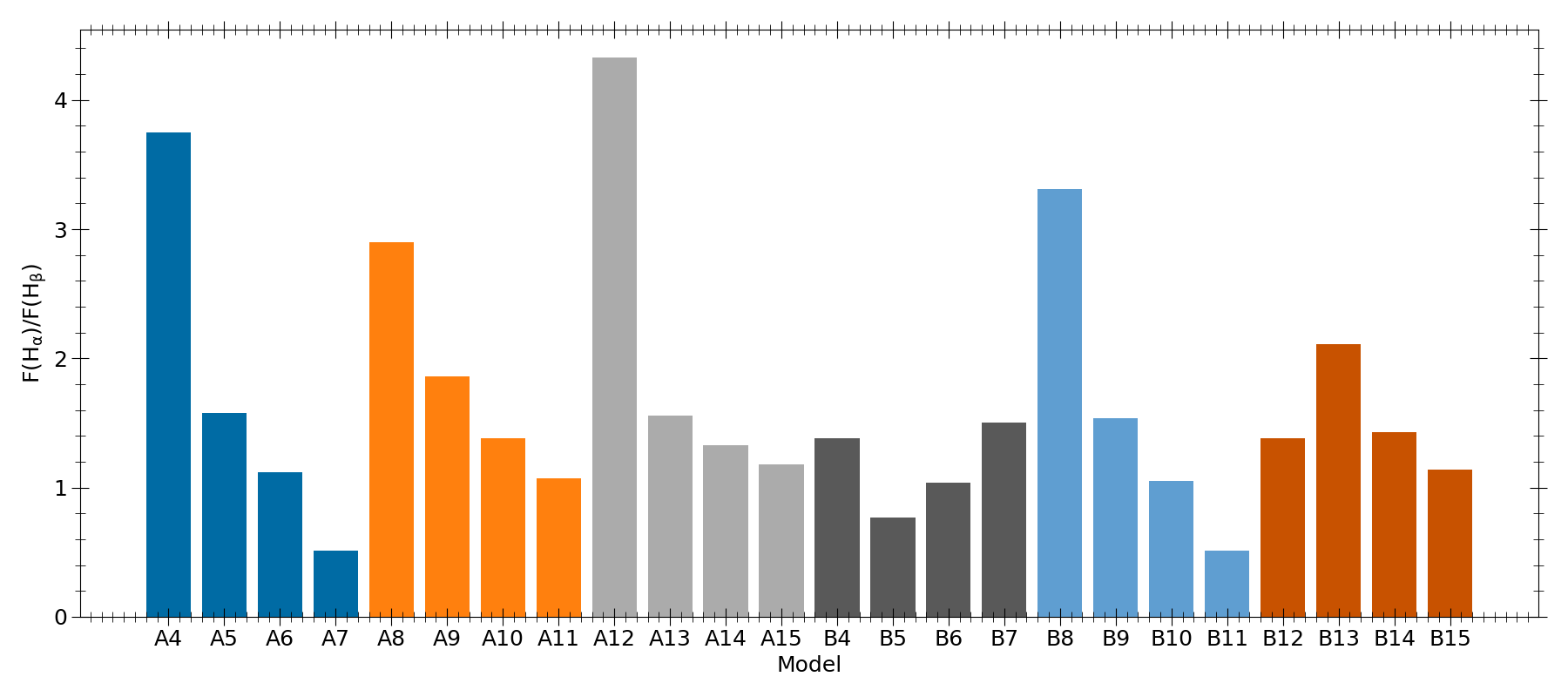}{0.85\textwidth}{(b)}}
\caption{The top panel shows F(\halpha) and the bottom panel shows the ratio of F(\halpha)/F(\hbeta) for the A-series (RSG) and B-series (BSG) models for the epoch closest to the maximum in the bolometric light curve. Each set of models with the same explosion energy with increasing CSM mass is shown in the same color. As seen in the figure, there is a correlation between F(\halpha) and the CSM mass, while the ratio of F(\halpha)/F(\hbeta) shows an anti-correlation. The properties of these lines for the models are listed in Tables \ref{tab:prop} \& \ref{tab:A_Series}. \label{fig:F_H_alpha_Lbolmax}}
\end{center}
\end{figure*}

Figure \ref{fig:F_H_alpha_Lbolmax} shows F(\halpha) in the upper panel and the ratio of F(\halpha)/F(\hbeta) in the lower panel, as calculated by the {\tt SuperLite} code at the epoch closest to the peak bolometric luminosity for the A- and B-series models. Unlike models with lower ejecta mass and energy, those with higher ejecta mass and explosion energy exhibit narrow helium emission lines. This is because of the higher gas temperatures and stronger photoionization in these models, even though the ratio of H/He is the same between the models with different ejecta masses. The spectra for C-series models look more or less featureless in the optical region. The models with higher CSM mass show broad He lines in the spectra, which form in the He-rich CSM envelope. The spectra seen in C-series models are consistent with the observations of the SLSN-I \citep{Nicholl:2016ab,Nicholl:2016ac,Quimby:2018aa}.

\subsection{Statistical Analysis of the Balmer series H-lines} \label{subsec:res_analysis}

In this section, we present the results of the data analysis performed on the properties of the Balmer series lines listed in Tables \ref{tab:A_Series} and \ref{tab:B_Series}. The purpose of this analysis is to find correlations between the H line predictions from {\tt SuperLite} code (hereafter, the targets) and the properties of the progenitor and the CSM (hereafter, the features) in a systematic manner. The properties listed in the table for H-lines include the amplitudes, fluxes, and line widths of the \halpha \ and the \hbeta \ lines, and the ratio of the amplitudes and the fluxes for the two lines. The dimensionality of the targets and the features is reduced by finding the correlations within each of the two sets using a statistical method called Variance Inflation Factor \citep[VIF,][]{Neter:1996applied} to detect the presence of multicollinearity within the features, and the same for the targets. It is a quantitative method that assesses the inflation of variance of the regression coefficient with other predictors. We use the Python package {\tt StatsModel}\footnote{\url{https://github.com/statsmodels/statsmodels}} for this method. A typical VIF score between 1.0 and 5.0 is acceptable. The features and targets that have scores higher than 10 are eliminated, in an iterative manner, starting with the highest scores, until the remaining features and targets have scores between 1.0 \& 5.0.
We find that the amplitudes, fluxes, and the line-widths of the H-lines are highly correlated. Thus, only the two uncorrelated quantities, the flux of \halpha \ line F(\halpha) and the ratio of the fluxes of the two lines, F(\halpha)/F(\hbeta), are considered as the targets for further analysis. Similarly, among the features, the uncorrelated quantities are the day since the peak luminosity, the explosion energy ($\rm E_{exp}$), the radius of the progenitor at the time of the explosion ($\rm R_{exp}$), the LTE/NLTE case, and the ratio of the ejecta mass to the that of the CSM ($\rm M_{ej}/M_{CSM}$).

After eliminating the highly correlated quantities, we employ the technique of a simple machine learning algorithm called K-Nearest Neighbors \citep[KNN,][]{cover:1967KNN}, where the targets are classified based on the proximity or the average value of their k nearest neighbors in the features space. The performance of our machine learning algorithm is evaluated using cross-validation techniques, where a part of the data is used for training the model and the rest is used to test its performance. The process is repeated until an optimized value of the mean squared error is achieved. We use the Python package SKLearn \citep{scikit-learn} for this analysis. 

\begin{figure}[ht!]
\begin{center}
\gridline{\fig{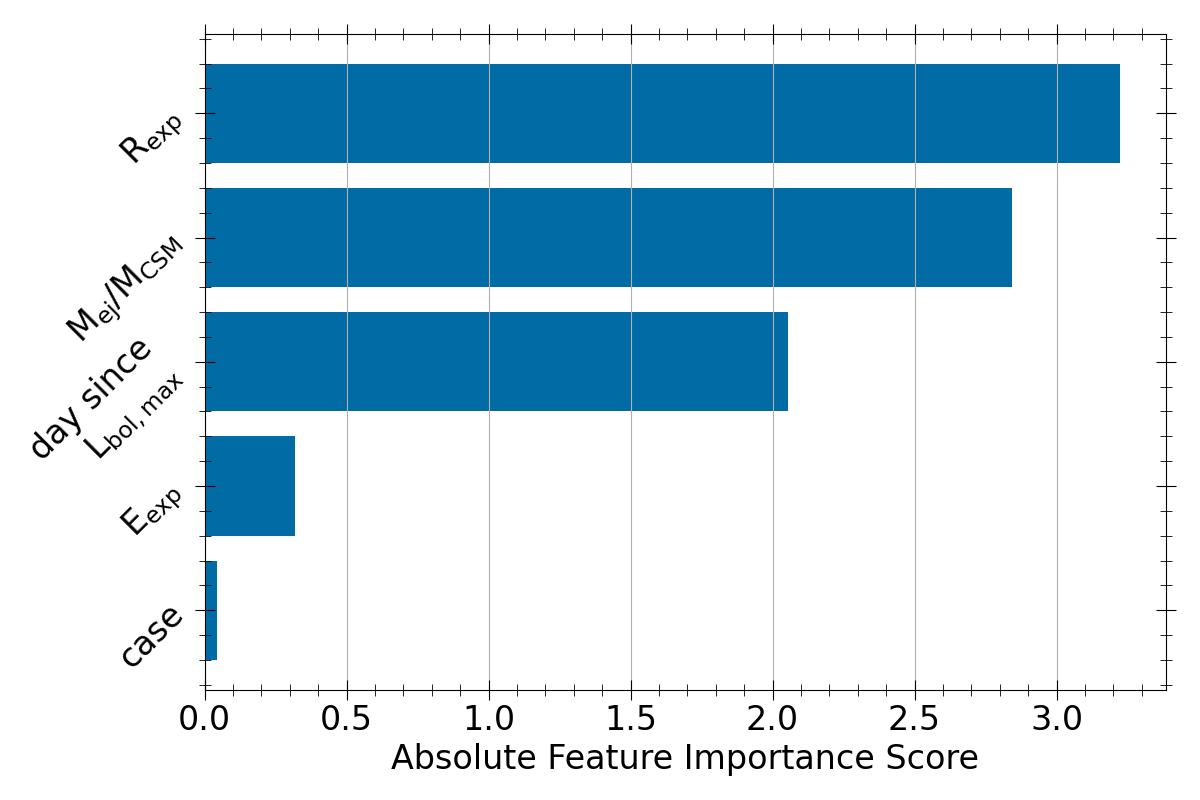}{0.45\textwidth}{(a)}}
\gridline{\fig{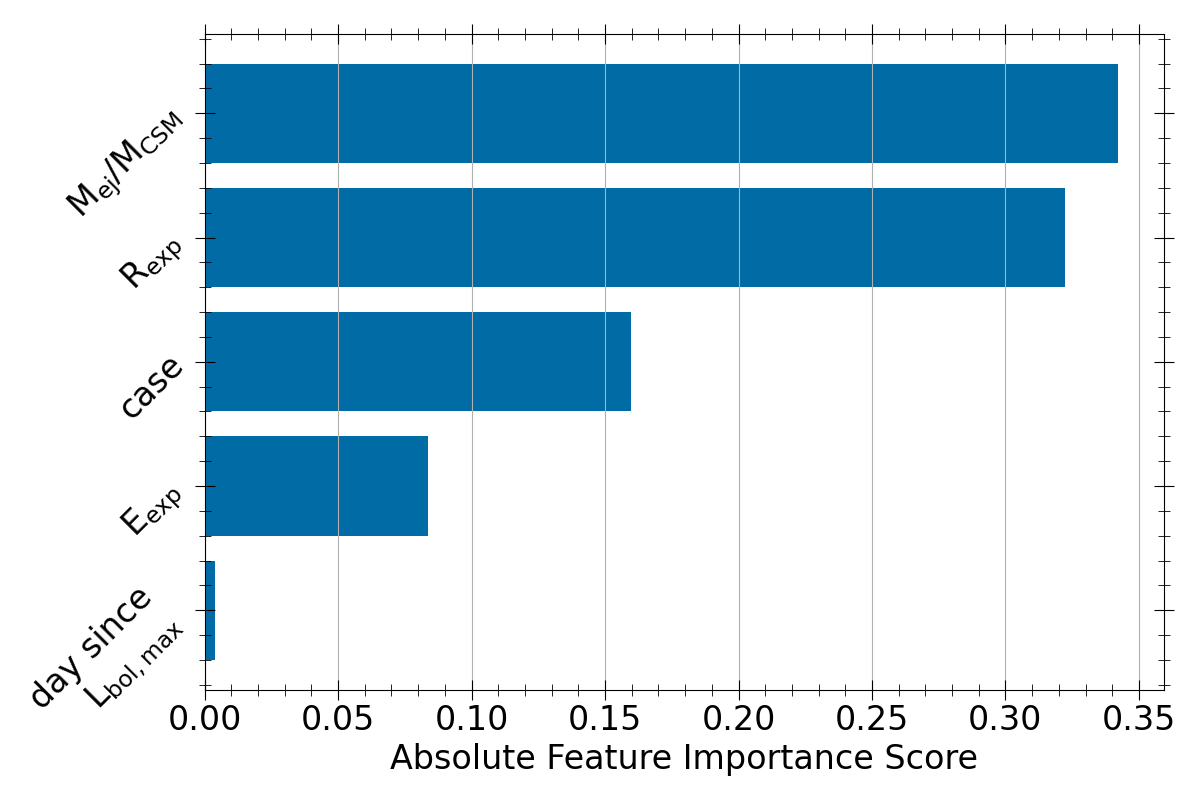}{0.45\textwidth}{(b)}}
\caption{The figure shows the ranked absolute feature importance scores for the K-nearest neighbor (KNN) analysis performed on the flux of the \halpha \ line, F(\halpha) in the panel (a), and the ratio F(\halpha)/F(\hbeta) in the panel (b).  \label{fig:KNN}}
\end{center}
\end{figure}

The results of the KNN analysis performed on the data for the targets and the features are displayed in Figure \ref{fig:KNN}. Panel (a) of the figure shows the ranked absolute feature importance scores for F(\halpha), and panel (b) shows the same for the ratio F(\halpha)/F(\hbeta). The feature importance score is an assessment of the model's performance following the perturbation of a specific feature. For a given model, the higher importance score implies that the said feature is more important in determining the value of the target. The scales for different models do not compare directly due to the differences in the model type, error metrics used, and data distribution. The KNN analysis for F(\halpha) gives the optimal k-value of 4 with root mean squared error (RMSE) of 0.33. The radius $\rm R_{exp}$ and the ratio $\rm M_{ej}/M_{CSM}$ are the most important features for the flux of the \halpha \ line. The other features are comparatively less significant, according to the analysis. The \halpha \ line flux correlates with $\rm R_{exp}$, while it anti-correlates with the ratio $\rm M_{ej}/M_{CSM}$. In the case of the ratio F(\halpha)/F(\hbeta), the optimal k-value is 2 with RMSE of 0.86. In this case, The radius $\rm R_{exp}$ again has the highest importance, followed by the LTE/NLTE case and the explosion energy $\rm E_{exp}$. The ratio of the fluxes is anti-correlated with $\rm R_{exp}$.

The combination of the dense CSM and higher velocity, higher mass ejecta leads to the formation of stronger, narrower emission lines of hydrogen due to higher densities and temperatures in the emitting region, which leads to more photoionization and in turn, more recombination as the gas cools down. As time progresses past the maximum luminosity, the photosphere recedes. The weaker recombination and the opacity effects reduce the flux in the \halpha line, reducing the ratio F(\halpha)/F(\hbeta). On the other hand, the lower velocity ejecta result in weaker, broader emission lines due to lower temperatures in the interaction region. It can be seen in tables \ref{tab:A_Series} \& \ref{tab:B_Series} that the line ratio gets stronger for the NLTE case, in comparison to the LTE case for calculation of the ionization states and excitation level population of hydrogen. It signifies that the NLTE conditions in the emitting region affect the \halpha \ line, making it more prominent. Both the \halpha \ line strength and the ratio F(\halpha)/F(\hbeta) are higher near the peak luminosity. The peak in the luminosity signifies the highest ionization due to stronger shock heating. As the shock moves outward and the density of the CSM decreases, the flux of the \halpha \ line and the ratio of the two lines decrease, and the lines become broader.

%\begin{figure*}[ht!]
%\begin{center}
%\gridline{\fig{ratio_ACN_A_Series.png}{0.45\textwidth}{(a)}
%         \fig{ratio_fluxes_A_Series.png}{0.45\textwidth}{(b)}}
%\caption{The ratio of continuum normalized amplitudes, A(\halpha)/A(\hbeta) is shown in the left panel and the ratio of fluxes, F(\halpha)/F(\hbeta) is shown in the left panel for A Series models. The values for the amplitudes and fluxes are noted in Table \ref{tab:A_Series}. Different marker symbols represent different CSM masses as noted in Table \ref{tab:prop}. Circles represent 0.2 \msun , squares 0.5 \msun , triangles 1.0 \msun , and diamonds 1.2 \msun .  \label{fig:ratio_A_Series}}
%\end{center}
%\end{figure*}
%
%
%\begin{figure*}[ht!]
%\begin{center}
%\gridline{\fig{ratio_ACN_B_Series.png}{0.45\textwidth}{(a)}
%         \fig{ratio_fluxes_B_Series.png}{0.45\textwidth}{(b)}}
%\caption{The ratio of continuum normalized amplitudes, A(\halpha)/A(\hbeta) is shown in the left panel and the ratio of fluxes, F(\halpha)/F(\hbeta) is shown in the left panel for B Series models. The values for the amplitudes and fluxes are noted in Table \ref{tab:A_Series}. Different marker symbols represent different CSM masses as noted in Table \ref{tab:prop}. Circles represent 0.2 \msun , squares 0.5 \msun , triangles 1.0 \msun , and diamonds 1.2 \msun . \label{fig:ratio_B_Series}}
%\end{center}
%\end{figure*}

%%%%% Duperfit evaluations
\subsection{Comparisons Against Observed Spectra}\label{subsec:duperfit_eval} %label if you need it at all

To compare the SuperLite models against observations, we utilized Duplicated Superfit in Python, or Duperfit \citep{Baer:2024ze}. This program follows the same algorithm as Superfit by D. Andrew Howell \citep{Howell:2005aa} to evaluate an input spectrum against observed template spectra. In essence, an observed spectrum at observed redshift $z$, $o(\lambda;z)$, is considered to be a combination of supernova and host galaxy light. While best practice dictates that host galaxy modeling should remove galaxy contamination in SN spectra, this is not always done for SNe, particularly those in hosts with no archival spectra.

%% Extra details on fitting added; in bold as detailed in your .txt file
Superfit assumes the intrinsic SN spectrum to match some archival template, $t(\lambda;z)$, along with a host galaxy, $g(\lambda;z)$, both at the same redshift. To further emulate host galaxy modeling, the algorithm considers the total extinction in the $V$ band, $A_V$, extinction parameter, $R_V \equiv A_V/E(B-V)$, and wavelength-specific extinction, $A(\lambda;R_V)$, given by a Cardelli law \citep{Cardelli:1989aa}, to define a relative reddening $r(\lambda;R_V) \equiv A(\lambda;R_V)/A_V$, for which $R_V$ is fixed to a user-defined value (3.1 by default) and $A_V$ is a free fitting parameter. The equation of the model to fit the observed spectrum, then, is
\begin{equation}
    o_\mathrm{mod}(\lambda;z) = Ct(\lambda;z)10^{-0.4A_Vr(\lambda;z)} + Dg(\lambda;z)
\end{equation}
\noindent for which $C$ and $D$ are free scaling parameters for fitting. The redshift $z$ may be either fixed or free as a grid-search parameter. This model is then fitted via a method of least squares, where weights may be equal or provided by inverse variance (from error flux), or a weight spectrum that removes known telluric lines, or with a user-provided weight file. For our fits, we instructed Duperfit to estimate error flux using an iterative B-spline fit and to apply the inverse variance.
%% end of extra details

To better evaluate the spectral type, Duperfit further utilizes a semi-automated scoring mechanism devised by \citet{Quimby:2018aa}, referred to as mean-index difference (MID) scoring within the program. The program returns matched templates in order from best to worst and assigns indexes $I$ to them from 0 to the total number of templates in that order. The program then computes a mean of the indices of the top 5 results of a given supernova type X, $\langle I_\mathrm{X}\rangle$, repeats the same process for a type Y, $\langle I_\mathrm{Y}\rangle$, then computes the difference:
\begin{equation}
    \Delta I_\mathrm{X-Y} = \langle I_\mathrm{X}\rangle - \langle I_\mathrm{Y}\rangle.
\end{equation}
where more negative values imply stronger agreement with type X, and more positive values imply stronger agreement with type Y. Values near zero suggest ambiguity.

To evaluate the best method for evaluating SuperLite spectra with Duperfit, we tested a successful Type Ia model generated by SuperLite to assess how well it could be recaptured by the program. We tested two different schemes for evaluating the model spectra. Across both schemes, we fitted the supernova template scaling factor ($0.01\leq C\leq 3$) and a relative extinction ($-4.5 \leq A_V\leq 2$) to account for any intrinsic reddening in specific supernova templates or color differences between SNe. The two schemes were then differentiated as follows:
\begin{itemize}[leftmargin=*]
    \item R1: Evaluate the spectrum at zero redshift with zero host galaxy emission modeled (i.e., the galaxy scaling factor $D$ was fixed to zero)
    \item R2: Artificially redshift the spectra to $z=0.1$, add a random amount of host galaxy emission to the input spectra via an Sc template, and set an Sc galaxy emission term to a free parameter (i.e., fit for $0\leq D\leq 3$), and perform a redshift grid-search with $0.03\leq z\leq 0.17$ with step size $\Delta z = 0.01$.
\end{itemize}
\begin{deluxetable*}{c|c|cccccc}
    \tablecaption{A table of MID scores, $\Delta I_\mathrm{X-Y}$, for the SN Ia model spectrum with the two sets of fitting parameters is provided. Rows represent type X, and columns represent type Y. Scores of identical types, $\Delta I_\mathrm{X-X}$, are intentionally left blank as they provide no information.}
    \tabletypesize{\footnotesize}
    \tablewidth{0pt}
    \tablehead{
        \colhead{} & \colhead{} & \colhead{SNIa} & \colhead{SNIb} & \colhead{SNIc} & \colhead{SNII} & \colhead{SNIIn} & \colhead{SLSN-I}
    }
    \startdata
        \multirow{6}{*}{R1} & SNIa & -- & -179.0 & -41.8 & 5.6 & 27.8 & -26.2\\
        & SNIb & 179.0 & -- & 137.2 & 184.6 & 206.8 & 152.8\\
        & SNIc & 41.8 & -137.2 & -- & 47.4 & 69.6 & 15.6\\
        & SNII & -5.6 & -184.6 & -47.4 & -- & 22.2 & -31.8\\
        & SNIIn & -27.8 & -206.8 & -69.6 & -22.2 & -- & -54.0 \\
        & SLSN-I & 26.2 & -152.8 & -15.6 & 31.8 & 54.0 & -- \\
        \tableline
        \multirow{6}{*}{R2} & SNIa & -- & -843.6 & -622.8 & -602.2 & -474.2 & -444.0\\
        & SNIb & 843.6 & -- & 220.8 & 241.4 & 369.4 & 399.6\\
        & SNIc & 622.8 & -220.8 & -- & 20.6 & 148.6 & 178.8\\
        & SNII & 602.2 & -241.4 & -20.6 & -- & 128.0 & 158.2\\
        & SNIIn & 474.2 & -369.4 & -148.6 & -128.0 & -- & 30.2\\
        & SLSN-I & 444.0 & -399.6 & -178.8 & -158.2 & -30.2 & -- \\
    \enddata
\end{deluxetable*}
\label{tab:w7-scores}

\begin{figure*}[ht!]
    \centering
    \gridline{\fig{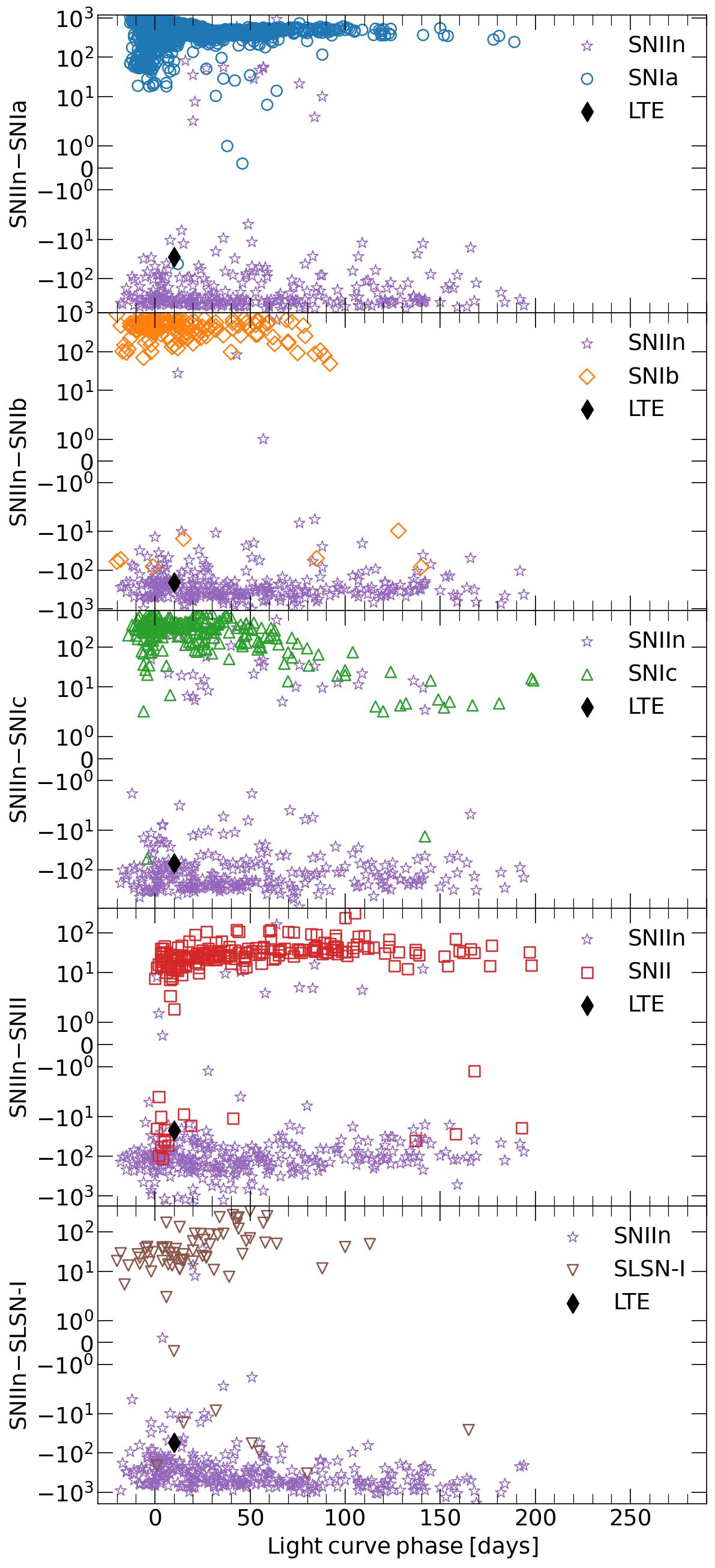}{0.45\textwidth}{(a)} %%% UPDATED FIG. WIDTH
              \fig{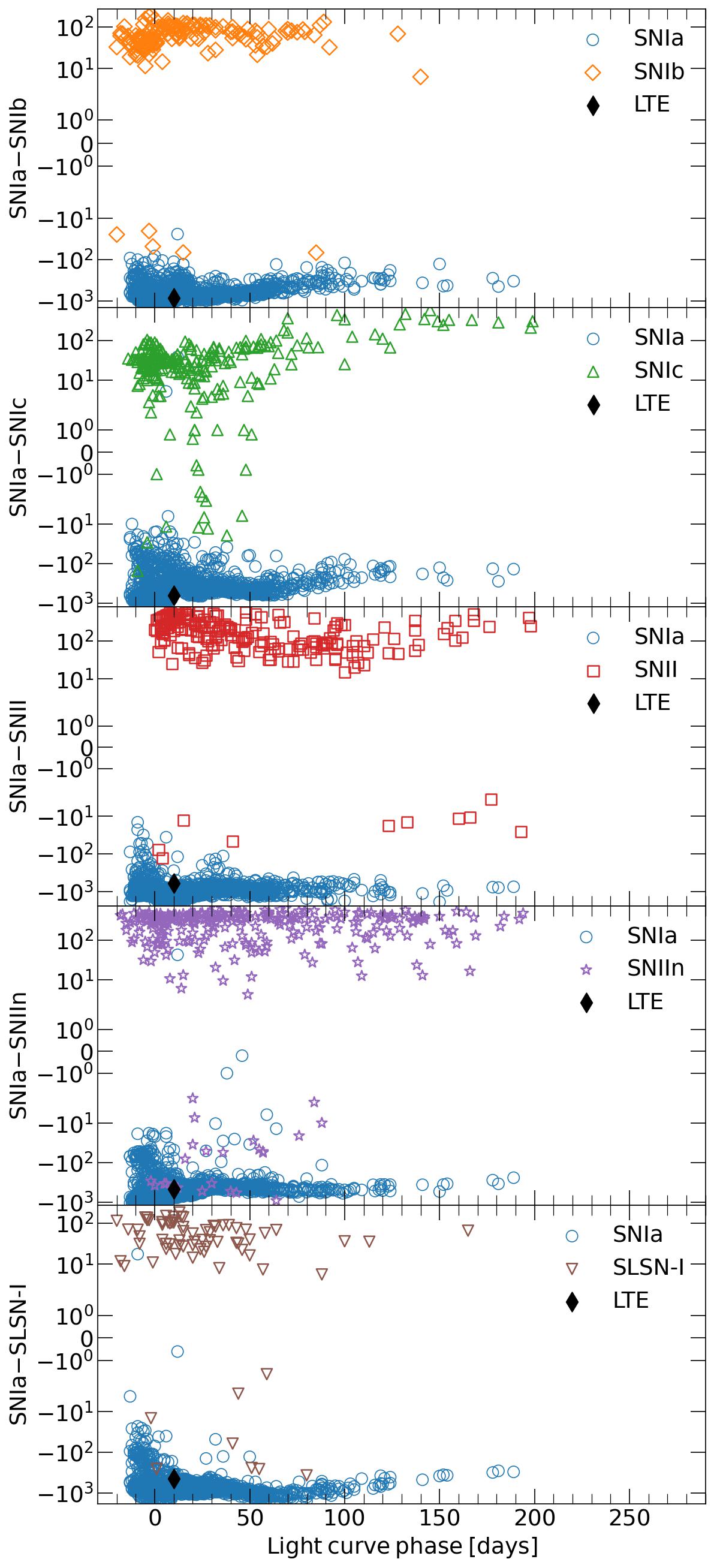}{0.45\textwidth}{(b)}}
    %% corrected caption
    \caption{MID score distributions for the SN Ia model fit are shown to visualize the scoring method. The fitting was performed to verify that the chosen fitting scheme reproduces a well-determined spectral type for a given model. Colored populations represent the template library's evaluations as a comparison set, while the black rhombus represents the input spectrum. The first type across the subplots represents the numerically best-matched type. Plots are symmetric-logarithmically scaled to clearly show the numerical separation between populations. (a) Distributions representing the fit with the R1 scheme. (b) Distributions representing the fit with the R2 scheme.}
    \label{fig:w7_df-comparison}
\end{figure*}

\noindent Evaluating these models in this way produces output from the program, as shown in Table \ref{tab:w7-scores}. Note that the off-diagonal elements are antisymmetric, so the ordering of scores is relatively inconsequential. Thus, the program may roughly determine a best-matched SN type by finding the maximum absolute median value of the rows or columns.

Numerically, we more readily recapture the model as an SNIa using the R2 parameters. However, to better highlight this distinction, we also plot the scores alongside the scoring of the template library. The template library contains default evaluations in Duperfit, where the code intentionally ignores matching spectra of the same supernova. These scores are then plotted in a symmetrical logarithmic scale to show the distinction between score populations. For these plots, the best-matched type for the input model is assigned type X, and all other types are assigned type Y. The comparison is visible in Fig. \ref{fig:w7_df-comparison}.

With the results from both tests, we determined that the R2 parameters provide the best approach for evaluating the SuperLite spectra. For each series, we chose models 1, 5, 9, and 13 to match explosion energy and noted the differences as CSM mass increased. We then used the MID scoring included in Duperfit to compute the best-fit empirical match of spectral type to these models.% The A-series evaluations are visible in Figs. \ref{fig:A1_5} \& \ref{fig:A9_13}, as an example.%, B-series in Figs. \ref{fig:B1_5} \& \ref{fig:B9_13}, and C-series in Figs. \ref{fig:C1_5} \& \ref{fig:C9_13}. To determine a best-matched type, we used the best match from Duperfit across the most phases. %%% Not bulking up the paper with these figures so much now

\begin{figure*}[!ht]
    \centering
    \gridline{\fig{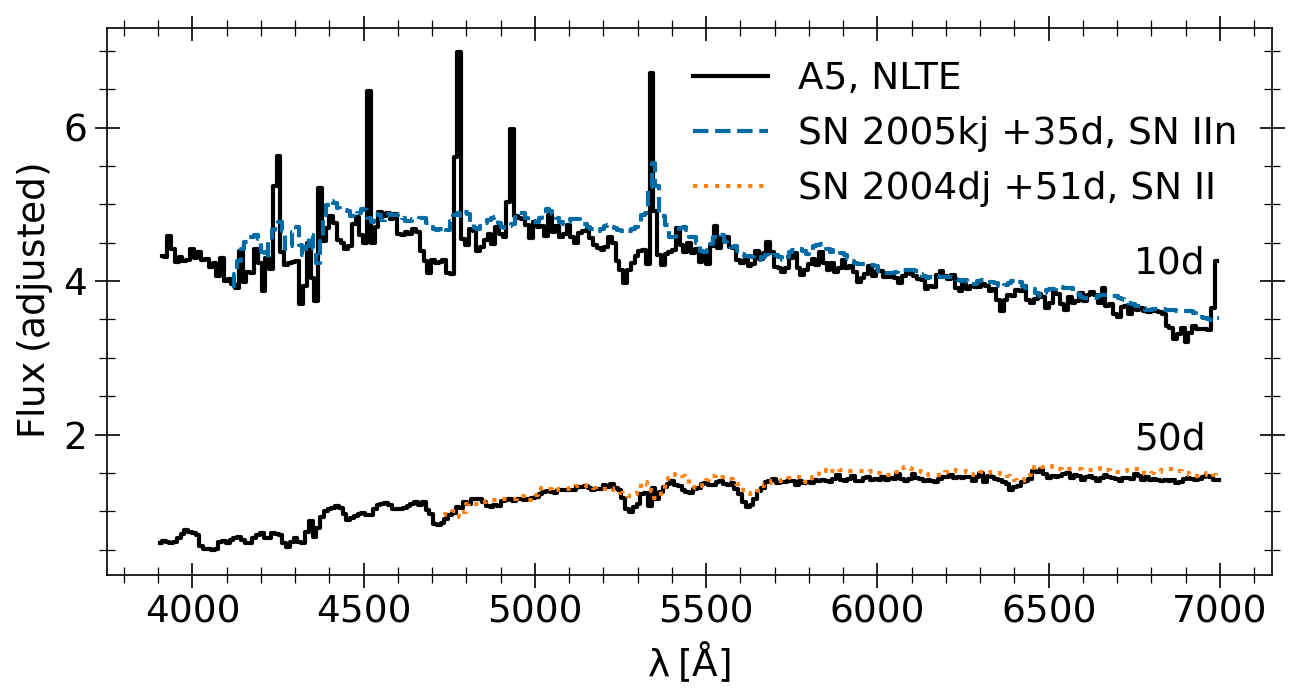}{0.45\linewidth}{(a)} %%% UPDATED FIG. WIDTH
              \fig{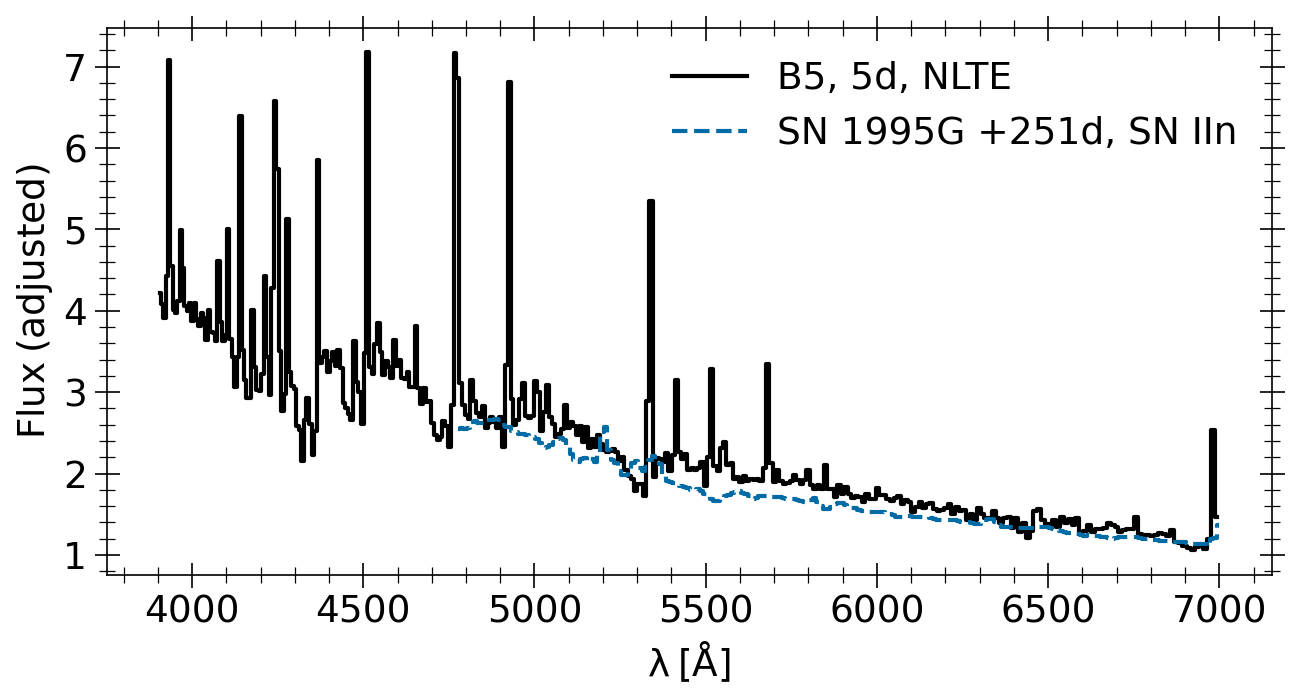}{0.45\linewidth}{(b)}}
    \gridline{\fig{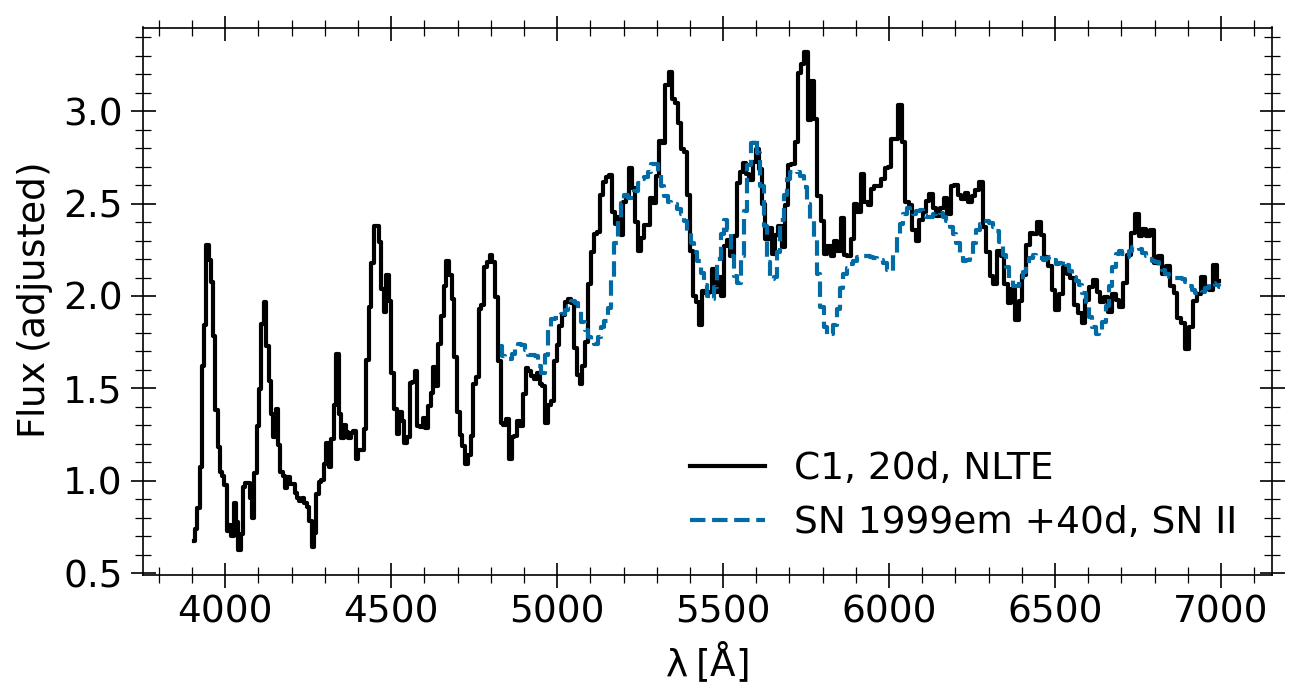}{0.45\linewidth}{(c)}
              \fig{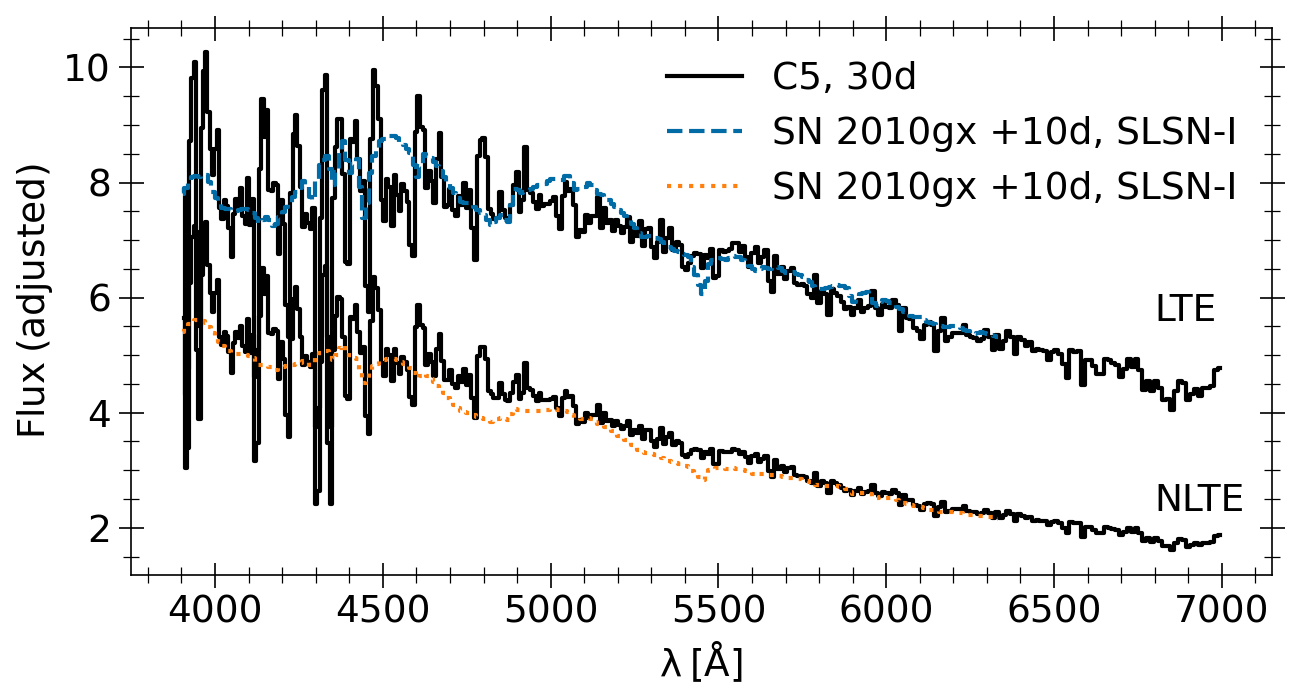}{0.45\linewidth}{(d)}}
    \caption{Plots of artificially reddened and galaxy-contaminated spectra for various models and the Duperfit-determined best-fit template model representing strong type correlation for the given phase. (a) The NLTE A5 model against Duperfit's best matches at 10 and 50 days, (b) the NLTE B5 model against its best match at 5 days, (c) the NLTE C1 model against Duperfit's best match at 20 days, and (d) the LTE C1 model against Duperfit's second-best match at 30 days and the NLTE C1 model against Duperfit's best match at 20 days.}
    \label{fig:df_spec-comparison}
\end{figure*}

The A1 model strongly agrees with non-interacting SNe II, while each increasing CSM mass model correlates more strongly with SNe IIn. The A5 model becomes more consistent with SNe II after 50 days from maximum light. We can see this evolution in A5 by comparing its best match at 10 days to 50 days past maximum, the Type IIn SN 2005kj and Type II SN 2004dj, respectively, in Fig. \ref{fig:df_spec-comparison}.

%%% A5 spectral comparison
%\begin{figure}[!ht]
%    \centering
    %\gridline{\fig{df_plot_A5_10d-nlte.jpg}{0.8\textwidth}{(a)}}         
    %\gridline{\fig{df_plot_A5_50d-nlte.jpg}{0.8\textwidth}{(b)}}
%    \includegraphics[width=\linewidth]{df_A5_matches.png} %%%%% New figure which takes less space while not sacrificing detail
%    \caption{Plot of the artificially reddened and galaxy-contaminated spectrum for the NLTE A5 model at 10 days and 50 days past maximum against Duperfit's best-matched evaluations for each phase.}
%    \label{fig:A5_10+50d-df}
%\end{figure}

The B1 model strongly correlates with SNe II, while the B5 model's best match was ambiguous between SNe II, IIn, and Ic. We can see in Fig. \ref{fig:df_spec-comparison} that, at the scored Ic-like phase 5 days past maximum light, the best-matched result remains the Type IIn SN 1995G with some strong convergence. The remaining B-series models are more consistent with SNe IIn up to the 50 days past maximum light in their evolutionary modeling.

%%% B5 spectral comparison
%\begin{figure}[ht!]
%    \centering
    %\includegraphics[width=\textwidth]{df_plot_B5_5d-nlte.jpg} %%%%% Replacing this figure
%    \includegraphics[width=\linewidth]{df_B5_match.png}
%    \caption{Plot of the artificially reddened and galaxy-contaminated spectrum for the NLTE B5 model at 5 days past maximum against Duperfit's best-matched evaluation for that phase.}
%    \label{fig:B5_5d-df}
%\end{figure}

The fitting of the C1 model was, curiously, ambiguous between SNe II, IIn, and Ic. This is likely due to Duperfit, like Superfit before it, numerically determining a best fit of the spectrum without regarding particular identifying features. In Fig. \ref{fig:df_spec-comparison}, we show the best-matched spectrum against the artificially reddened and galaxy-contaminated spectrum, using the fitted parameters. While the general shape has converged to a good fit, the features do not necessarily align.

The C5 model showed consistency with hydrogen-deficient SLSNe, though its early phases' top results did not converge to satisfactory goodness-of-fit values. However, 30 days past maximum light, the constraint improves significantly, as shown in Fig. \ref{fig:df_spec-comparison}, where the second-best match for the LTE model and the best match for the NLTE model closely resemble the Type I SLSN 2010gx at 10 days past maximum light. Note, in particular, the similarity of the LTE model continuum to the O II absorption features in the 4000-4700\AA\space range, which \citet{Quimby:2018aa} identified as unique characteristic distinguishing early-phase SLSNe-I. However, the poor physical representation of CSM interaction in LTE models means that this match should be treated with some scrutiny.

Finally, C9 and C13 had the best fits with SNe IIn in their scores. While the C-series models lacked H lines in their spectra, Superfit (and, by extension, Duperfit) does not weigh line presence more heavily than other features when evaluating the best fit. Furthermore, the template library of SNe IIn is fairly heterogeneous, including events of weaker H-line strength. For instance, the C9 model at 30 days had a best match with the Type IIn SN 2008fq, which itself had fairly weak lines. Finally, some higher-ranked matches for certain spectra had a higher-scaled galaxy template over the SN template, which may skew that result.

%%% C1 spectral comparison
%\begin{figure}[ht!]
%    \centering
    %\includegraphics[width=\textwidth]{df_plot_C1_20d-nlte.jpg}    %%%%% Replacing this figure
%    \includegraphics[width=\linewidth]{df_C1_match.png}
%    \caption{Plot of the artificially reddened and galaxy-contaminated spectrum for the NLTE C1 model at 20 days past maximum against Duperfit's best-matched evaluation for that phase.}
%    \label{fig:C1_20d-df}
%\end{figure}

%%% C5 spectral comparison
%\begin{figure}[!ht]
%    \centering
    %\gridline{\fig{df_plot_C5_30d-lte.jpg}{\textwidth}{(a)}}         %%%% Replacing subfigures soon
    %\gridline{\fig{df_plot_C5_30d-nlte.jpg}{\textwidth}{(b)}}
%    \includegraphics[width=\linewidth]{df_C5_matches.png}
%    \caption{Plot of the artificially reddened and galaxy-contaminated spectrum for the LTE and NLTE C5 model at 30 days past maximum against Duperfit's second (for LTE) and first (for NLTE) best-matched evaluation.}
%    \label{fig:C5_30d-df}
%\end{figure}

%%% NEW: All spectral comparisons

\section{Conclusions} \label{sec:conclusions}

In this study, we explored the question of whether the observed diversity in supernovae (SNe) and superluminous supernovae (SLSNe) can be attributed to the diversity in the circumstellar material (CSM) surrounding massive stars. Our analysis was based on three model series -- A, B, and C -- each associated with different types of single star progenitors: a red supergiant (RSG), a yellow supergiant (YSG), and a blue supergiant (BSG). These progenitors were evolved using the {\tt MESA} code, and the subsequent SN ejecta were simulated using the hydrodynamic code {\tt STELLA}. The evolution profiles generated by {\tt STELLA} were further processed with the Monte Carlo radiation transport code {\tt SuperLite} to derive the spectral evolution at various epochs.

\begin{itemize}[leftmargin=*]
    \item \textbf{Influence of CSM Mass and Composition:} We found that the mass and composition of the CSM are critical in determining the type of SN, as expected. Specifically, our models suggest that the presence and richness of the CSM influence both the spectral class and light curve characteristics of the resultant transient, as well as its specific spectral evolution, particularly in terms of the relative strengths of the \halpha \ and \hbeta \ lines.
    \item \textbf{Spectral Diversity Across Models:}
    \begin{itemize}
        \item Models with RSG and YSG progenitors typically resembled SN-IIP in the absence of a CSM envelope but transitioned to SN-IIn and SLSN-II profiles when an H-rich CSM was present.
        \item BSG progenitors, due to their compact nature, predominantly produced light curves and spectra similar to SN-Ib/c. In scenarios with a He-rich CSM, the models predicted brighter light curves, characteristic of some SLSN-I events.
    \end{itemize}
    \item \textbf{Line Strength and Flux Ratios:} Our analysis highlighted a correlation between the progenitor properties (such as radius at explosion) and the CSM characteristics (e.g., mass ratio of ejecta to CSM) with the observed strengths and ratios of Balmer series lines, particularly \halpha \ and \hbeta. Notably:
    \begin{itemize}
        \item The strength of \halpha \ was significantly greater in models with larger progenitor radii and thicker H-rich CSM layers, indicating a stronger interaction and greater hydrogen ionization.
        \item Conversely, \hbeta \ was relatively stronger in configurations with higher explosion energies and less dense CSM, suggesting different excitation conditions.
    \end{itemize}
\end{itemize}

Our findings provide valuable insights into how the physical characteristics of progenitors and their surrounding CSM contribute to the spectroscopic evolution of these transients. The strength and ratio of Balmer lines, as influenced by various model parameters, offer a promising method for inferring progenitor and CSM properties from observed spectra.

Looking ahead, we aim to extend this research to hydrogen-poor events, exploring a broader range of parameters, such as different CSM shapes and geometries, using advanced simulations with SuperLite. This expansion could potentially unveil further nuances in the spectroscopic signatures of luminous transients, enhancing our understanding of their progenitors and explosion mechanisms.

\begin{acknowledgments}
We would like to thank Ryan Wollaeger, J. Craig Wheeler, Chris Fryer and Nathan Smith for useful discussions. G.W would like to thank Michael Attia for his valuable inputs on the ML techniques for data analysis. G.W. and E.C. would like to thank the National Science Foundation (NSF) for their support made possible by the NSF grant AST-1907617. EC would also like to thank NASA and the Smithsonian Astrophysical Observatory (SAO) for their support via the Chandra X-ray Observatory (CXO) theory grant TM4-25003X. 
\end{acknowledgments}

%% To help institutions obtain information on the effectiveness of their 
%% telescopes the AAS Journals has created a group of keywords for telescope 
%% facilities.
%
%% Following the acknowledgments section, use the following syntax and the
%% \facility{} or \facilities{} macros to list the keywords of facilities used 
%% in the research for the paper.  Each keyword is check against the master 
%% list during copy editing.  Individual instruments can be provided in 
%% parentheses, after the keyword, but they are not verified.

%\vspace{5mm}
%\facilities{HST(STIS), Swift(XRT and UVOT)}

%% Similar to \facility{}, there is the optional \software command to allow 
%% authors a place to specify which programs were used during the creation of 
%% the manuscript. Authors should list each code and include either a
%% citation or url to the code inside ()s when available.

\software{{\tt SuperNu}\citep{Wollaeger:2013aa,Wollaeger:2014aa}, 
          {\tt SuperLite} \citep{Wagle:2023aa,Wagle:2023ze}, 
          {\it Duperfit} \citep{Howell:2005aa,Baer:2024ze}, 
          {\tt STELLA} \citep{Blinnikov:1993aa,Blinnikov:1998aa,Blinnikov:2004aa,Blinnikov:2006aa}, 
          {\tt MESA} r-15140 \citep{Paxton:2011aa,Paxton:2013aa,Paxton:2015aa,Paxton:2018aa,Paxton:2019aa}, 
          {\it NumPy} \citep{2020NumPy-Array}, 
          {\it Pandas} \citep{McKinney:2010data}, 
          {\it Astropy} \citep{Astropy-Collaboration:2022aa, Price:2018astropy},
          {\it Specutils} \citep{specutil_2022},
          {\it SKLearn} {\citep{scikit-learn}}
          }
          
\appendix

\section{Appendix A}

\setcounter{table}{0}
\renewcommand{\thetable}{A\arabic{table}}

\begin{longrotatetable}
\begin{deluxetable*}{cccccccccccccc}
\tablecaption{Hydrogen Line Properties for A Series Models. \label{tab:A_Series}}
\tabletypesize{\footnotesize}
\tablewidth{0pt}
\tablehead{
\colhead{Model} & \colhead{day since} & \colhead{$\rm L_{bol}$} & \colhead{H-Opacity} & \colhead{A(\halpha)$_{CN}$} & \colhead{A(\halpha)$_{CS}$} & \colhead{F(\halpha)} & \colhead{EW(\halpha)} & \colhead{A(\hbeta)$_{CN}$} & \colhead{A(\halpha)$_{CS}$} & \colhead{F(\hbeta)} & \colhead{EW(\hbeta)} & \colhead{A(\halpha)$_{CN}$/} & \colhead{F(\halpha)/} \\ [-.2cm]
\colhead{} & \colhead{$\rm L_{bol,max}$} & \colhead{$\rm (\times10^{42})$} & \colhead{} & \colhead{} & \colhead{$\rm (\times10^{-2})$} & \colhead{$\rm (\times10^{-1})$} & \colhead{} & \colhead{} & \colhead{$\rm (\times10^{-2})$} & \colhead{$\rm (\times10^{-1})$} & \colhead{} & \colhead{A(\halpha)$_{CN}$} & \colhead{F(\hbeta)}
}
\decimalcolnumbers
\startdata
\multirow{8}{*}{A4} & \multirow{2}{*}{0.1} & \multirow{2}{*}{5.82} & LTE & 2.82 & 6.61 & 9.66$\pm$1.37 & -26.59 & 1.52 & 2.87 & 2.58$\pm$1.17 & -4.63 & 1.86 & 3.75 \\
   & & & NLTE & 2.50 & 5.47 & 7.68$\pm$1.24 & -21.09 & 1.55 & 3.07 & 2.68$\pm$1.18 & -4.83 & 1.61 & 2.87 \\ 
   \cline{2-14}
   & \multirow{2}{*}{5} & \multirow{2}{*}{4.18} & LTE & 2.47 & 4.38 & 6.69$\pm$1.02 & -22.50 & 1.73 & 3.35 & 3.42$\pm$1.04 & -7.44 & 1.43 & 1.95 \\
   & & & NLTE & 2.51 & 4.48 & 6.29$\pm$1.01 & -21.17 & 1.69 & 3.16 & 3.30$\pm$1.03 & -7.18 & 1.49 & 1.91 \\
   \cline{2-14}
   & \multirow{2}{*}{10} & \multirow{2}{*}{3.09} & LTE & 1.76 & 1.81 & 2.45$\pm$0.62 & -10.32 & 1.68 & 2.51 & 2.40$\pm$0.82 & -6.47 & 1.05 & 1.02 \\
   & & & NLTE & 1.96 & 2.30 & 3.22$\pm$0.68 & -13.52 & 1.65 & 2.44 & 2.50$\pm$0.83 & -6.71 & 1.19 & 1.29 \\
   \cline{2-14}
   & \multirow{2}{*}{20} & \multirow{2}{*}{1.94} & LTE & 1.42 & 0.66 & 0.92$\pm$0.36 & -5.85 & 1.43 & 1.08 & 0.80$\pm$0.52 & -3.15 & 1.00 & 1.15 \\
   & & & NLTE & 1.45 & 0.71 & 0.93$\pm$0.37 & -5.88 & 1.36 & 0.90 & 0.58$\pm$0.51 & -2.29 & 1.07 & 1.60 \\
   \cline{2-14}
\tableline
\multirow{8}{*}{A5} & \multirow{2}{*}{0.1} & \multirow{2}{*}{9.36} & LTE & 3.86 & 15.3 & 24.1$\pm$2.70 & -45.24 & 2.75 & 15.5 & 15.3$\pm$2.59 & -17.20 & 1.41 & 1.58 \\
   & & & NLTE & 3.80 & 14.8 & 22.6$\pm$2.61 & -42.64 & 2.67 & 14.8 & 14.2$\pm$2.53 & -15.95 & 1.42 & 1.60 \\ 
   \cline{2-14}
   & \multirow{2}{*}{5} & \multirow{2}{*}{7.28} & LTE & 2.56 & 7.39 & 11.8$\pm$1.70 & -24.84 & 2.73 & 13.5 & 12.0$\pm$2.23 & -15.32 & 0.94 & 0.99 \\
   & & & NLTE & 2.84 & 8.73 & 13.7$\pm$1.84 & -28.90 & 2.70 & 13.2 & 11.8$\pm$2.20 & -15.28 & 1.05 & 1.16 \\  
   \cline{2-14}
   & \multirow{2}{*}{10} & \multirow{2}{*}{5.34} & LTE & 1.90 & 3.55 & 4.71$\pm$1.08 & -11.96 & 2.37 & 8.68 & 7.51$\pm$1.65 & -11.86 & 0.80 & 0.63 \\
   & & & NLTE & 2.07 & 4.19 & 5.73$\pm$1.15 & -14.61 & 2.34 & 8.44 & 8.28$\pm$1.67 & -13.10 & 0.89 & 0.69 \\ 
   \cline{2-14}
   & \multirow{2}{*}{20} & \multirow{2}{*}{3.26} & LTE & 1.65 & 1.82 & 1.98$\pm$0.68 & -7.09 & 1.48 & 2.03 & 1.60$\pm$0.87 & -3.79 & 1.11 & 1.24 \\
   & & & NLTE & 1.68 & 1.89 & 2.16$\pm$0.69 & -7.77 & 1.45 & 1.91 & 1.65$\pm$0.87 & -3.93 & 1.16 & 1.31 \\ 
\tableline
\multirow{6}{*}{A6} & \multirow{2}{*}{0.1} & \multirow{2}{*}{14.3} & LTE & 5.53 & 29.4 & 49.8$\pm$4.64 & -76.60 & 4.51 & 43.3 & 44.4$\pm$5.32 & -35.90 & 1.23 & 1.12 \\
   & & & NLTE & 5.57 & 29.9 & 51.6$\pm$4.74 & -78.85 & 4.38 & 41.8 & 42/7$\pm$5.20 & -34.61 & 1.27 & 1.21 \\ 
   \cline{2-14}
   % & \multirow{2}{*}{5} & \multirow{2}{*}{} & LTE &  &  & $\pm$ &  &  &  & $\pm$ &  &  & \\
   % & & & NLTE &  &  & $\pm$ &  &  &  & $\pm$ &  &  &  \\  
   % \cline{2-14}
   & \multirow{2}{*}{10} & \multirow{2}{*}{9.08} & LTE & 2.82 & 10.4 & 16.6$\pm$2.21 & -29.22 & 3.13 & 21.2 & 21.2$\pm$2.33 & -21.23 & 0.90 & 0.79 \\
   & & & NLTE & 2.87 & 10.7 & 17.1$\pm$2.24 & -29.91 & 3.06 & 20.4 & 21.4$\pm$3.17 & -21.59 & 0.94 & 0.80 \\ 
   \cline{2-14}
   & \multirow{2}{*}{20} & \multirow{2}{*}{5.78} & LTE & 2.14 & 5.17 & 7.28$\pm$1.37 & -16.09 & 1.75 & 5.59 & 4.24$\pm$1.65 & -5.69 & 1.23 & 1.72 \\
   & & & NLTE & 2.2 & 5.39 & 7.38$\pm$1.39 & -16.41 & 1.78 & 5.83 & 4.21$\pm$1.66 & -5.64 & 1.23 & 1.75 \\ 
\tablebreak
\multirow{8}{*}{A7} & \multirow{2}{*}{0.1} & \multirow{2}{*}{19.9} & LTE & 3.75 & 17.3 & 42.5$\pm$4.02 & -58.49 & 5.95 & 77.9 & 83.3$\pm$8.81 & -52.97 & 0.63 & 0.51 \\
   & & & NLTE & 7.67 & 48.8 & 87.3$\pm$7.24 & -119.05 & 5.77 & 74.7 & 79.4$\pm$8.51 & -50.66 & 1.33 & 1.10 \\ 
   \cline{2-14}
   & \multirow{2}{*}{5} & \multirow{2}{*}{17.6} & LTE & 2.77 & 11.3 & 28.2$\pm$3.25 & -36.23 & 4.45 & 54.8 & 59.1$\pm$6.96 & -37.14 & 0.62 & 0.48 \\
   & & & NLTE & 3.83 & 22.0 & 43.2$\pm$4.25 & -54.48 & 4.41 & 54.3 & 57.4$\pm$6.89 & -36.03 & 0.87 & 0.75 \\  
   \cline{2-14}
   & \multirow{2}{*}{10} & \multirow{2}{*}{14.4} & LTE & 2.25 & 85.4 & 18.1$\pm$2.57 & -24.24 & 3.33 & 33.2 & 34.6$\pm$4.91 & -24.21 & 0.68 & 0.52 \\   
   & & & NLTE & 3.84 & 21.4 & 39.1$\pm$3.98 & -51.85 & 3.26 & 32.6 & 34.5$\pm$4.89 & -23.98 & 1.18 & 1.13 \\ 
   \cline{2-14}
   & \multirow{2}{*}{20} & \multirow{2}{*}{9.71} & LTE & 2.70 & 10.8 & 17.9$\pm$2.40 & -28.17 & 2.07 & 12.5 & 12.3$\pm$2.91 & -10.49 & 1.31 & 1.46 \\
   & & & NLTE & 2.85 & 11.8 & 19.4$\pm$2.52 & -20.35 & 2.14 & 13.4 & 11.5$\pm$2.92 & -9.80 & 1.33 & 1.69 \\
\tableline
\multirow{8}{*}{A8} & \multirow{2}{*}{0.1} & \multirow{2}{*}{7.35} & LTE & 2.05 & 5.09 & 6.98$\pm$1.41 & -14.49 & 1.31 & 2.24 & 2.41$\pm$1.47 & -3.31 & 1.57 & 2.90 \\
   & & & NLTE & 1.83 & 4.00 & 4.99$\pm$1.28 & -10.32 & 1.29 & 2.14 & 2.11$\pm$1.46 & -2.90 & 1.41 & 2.36 \\ 
   \cline{2-14}
   & \multirow{2}{*}{5} & \multirow{2}{*}{5.04} & LTE & 2.05 & 3.92 & 5.50$\pm$1.10 & -14.66 & 1.45 & 2.49 & 2.16$\pm$1.14 & -3.91 & 1.41 & 2.54 \\
   & & & NLTE & 1.97 & 3.63 & 4.86$\pm$1.06 & -12.93 & 1.43 & 2.39 & 1.88$\pm$1.13 & -3.42 & 1.37 & 2.58 \\  
   \cline{2-14}
   & \multirow{2}{*}{10} & \multirow{2}{*}{3.58} & LTE & 1.60 & 1.71 & 2.81$\pm$0.73 & -9.77 & 1.55 & 2.22 & 2.04$\pm$0.86 & -5.02 & 1.03 & 1.38 \\
   & & & NLTE & 1.66 & 1.89 & 2.79$\pm$0.74 & -9.74 & 1.50 & 2.02 & 1.68$\pm$0.84 & -4.14 & 1.11 & 1.67  \\ 
   \cline{2-14}
   & \multirow{2}{*}{20} & \multirow{2}{*}{2.27} & LTE & 1.31 & 0.58 & 0.50$\pm$0.40 & -2.65 & 1.39 & 1.06 & 0.97$\pm$0.55 & -3.57 & 0.94 & 0.52 \\
   & & & NLTE & 1.39 & 0.73 & 0.68$\pm$0.41 & -3.60 & 1.32 & 0.87 & 0.56$\pm$0.53 & -2.08 & 1.05 & 1.20 \\ 
\tableline
\multirow{8}{*}{A9} & \multirow{2}{*}{0.1} & \multirow{2}{*}{12.6} & LTE & 2.63 & 12.3 & 19.3$\pm$2.75 & -25.71 & 1.92 & 11.3 & 10.4$\pm$2.89 & -8.40 & 1.37 & 1.86 \\
   & & & NLTE & 2.57 & 11.9 & 18.1$\pm$2.68 & -23.98 & 1.77 & 9.48 & 8.60$\pm$2.78 & -6.97 & 1.46 & 2.11 \\ 
   \cline{2-14}
   & \multirow{2}{*}{5} & \multirow{2}{*}{9.12} & LTE & 2.35 & 8.42 & 12.8$\pm$2.06 & -20.46 & 2.06 & 10.4 & 9.70$\pm$2.39 & -9.88 & 1.14 & 1.31 \\
   & & & NLTE & 2.49 & 9.25 & 13.4$\pm$2.12 & -21.60 & 2.07 & 10.5 & 10.2$\pm$2.41 & -10.37 & 1.20 & 1.32 \\  
   \cline{2-14}
   & \multirow{2}{*}{10} & \multirow{2}{*}{6.49} & LTE & 1.75 & 3.72 & 5.71$\pm$1.32 & -11.57 & 2.11 & 8.37 & 7.20$\pm$1.85 & -9.52 & 0.83 & 0.79 \\
   & & & NLTE & 1.93 & 4.57 & 7.36$\pm$1.43 & -14.94 & 2.09 & 8.18 & 7.55$\pm$1.84 & -10.07 & 0.92 & 0.98 \\ 
   \cline{2-14}
   & \multirow{2}{*}{20} & \multirow{2}{*}{4.07} & LTE & 1.40 & 1.42 & 1.64$\pm$0.79 & -4.67 & 1.51 & 2.57 & 1.48$\pm$0.10 & -2.97 & 0.93 & 1.11 \\
   & & & NLTE & 1.46 & 1.59 & 1.63$\pm$0.71 & -4.67 & 1.44 & 2.20 & 1.41$\pm$1.01 & -2.81 & 1.01 & 1.16 \\ 
\tablebreak
\multirow{8}{*}{A10} & \multirow{2}{*}{0.1} & \multirow{2}{*}{19.3} & LTE & 4.09 & 27.8 & 46.1$\pm$4.87 & -51.11 & 3.02 & 34.1 & 33.5$\pm$5.27 & -19.87 & 1.35 & 1.38 \\
   & & & NLTE & 4.03 & 27.3 & 45.7$\pm$4.83 & -50.74 & 2.96 & 32.8 & 32.6$\pm$5.17 & -19.50 & 1.36 & 1.40 \\ 
   \cline{2-14}
   & \multirow{2}{*}{5} & \multirow{2}{*}{15.7} & LTE & 3.32 & 20.6 & 35.5$\pm$4.04 & -40.03 & 2.85 & 29.0 & 30.9$\pm$4.78 & -19.72 & 1.16 & 1.15 \\
   & & & NLTE & 3.53 & 22.5 & 36.8$\pm$4.19 & -41.53 & 2.85 & 28.9 & 28.8$\pm$4.71 & -18.43 & 1.24 & 1.28 \\  
   \cline{2-14}
   & \multirow{2}{*}{10} & \multirow{2}{*}{11.1} & LTE & 2.55 & 11.2 & 18.1$\pm$2.59 & -25.05 & 2.78 & 21.9 & 22.5$\pm$3.68 & -18.28 & 0.92 & 0.80 \\
   & & & NLTE & 2.65 & 11.9 & 19.2$\pm$2.66 & -26.61 & 2.66 & 20.4 & 20.5$\pm$3.55 & -16.69 & 1.00 & 0.94 \\ 
   \cline{2-14}
   & \multirow{2}{*}{20} & \multirow{2}{*}{6.96} & LTE & 1.63 & 3.47 & 4.19$\pm$1.36 & -7.59 & 1.74 & 6.43 & 4.75$\pm$1.90 & -5.50 & 0.93 & 0.88 \\
   & & & NLTE & 1.77 & 4.24 & 5.39$\pm$1.44 & -9.79 & 1.64 & 5.58 & 4.30$\pm$1.87 & -4.93 & 1.08 & 1.25 \\
\tableline
\multirow{8}{*}{A11} & \multirow{2}{*}{0.1} & \multirow{2}{*}{27.8} & LTE & 6.18 & 51.1 & 91.2$\pm$7.95 & -92.34 & 4.95 & 81.9 & 85.3$\pm$9.76 & -41.13 & 1.25 & 1.07 \\
   & & & NLTE & 6.21 & 51.5 & 94.5$\pm$8.10 & -95.22 & 4.58 & 7.40 & 74.8$\pm$9.20 & -37.91 & 1.36 & 1.20 \\ 
   \cline{2-14}
   & \multirow{2}{*}{5} & \multirow{2}{*}{-24.6} & LTE & 3.33 & 24.2 & 46.9$\pm$5.00 & -44.29 & 4.31 & 71.0 & 75.1$\pm$9.04 & -34.99 & 0.77 & 0.62 \\
   & & & NLTE & 5.00 & 42.6 & 78.3$\pm$7.14 & -73.39 & 4.12 & 66.4 & 70.3$\pm$8.64 & -33.00 & 1.21 & 1.11 \\ 
   \cline{2-14}
   & \multirow{2}{*}{10} & \multirow{2}{*}{19.1} & LTE & 2.34 & 12.7 & 25.5$\pm$3.45 & -26.42 & 3.47 & 46.7 & 50.8$\pm$6.79 & -26.87 & 0.67 & 0.50 \\
   & & & NLTE & 4.15 & 30.4 & 55.2$\pm$5.45 & -57.12 & 3.61 & 49.3 & 53.0$\pm$6.97 & -28.07 & 1.15 & 1.04 \\ 
   \cline{2-14}
   & \multirow{2}{*}{20} & \multirow{2}{*}{11.9} & LTE & 2.15 & 9.02 & 15.1$\pm$2.49 & -19.32 & 1.92 & 12.7 & 11.4$\pm$3.27 & -8.27 & 1.12 & 1.32 \\
   & & & NLTE & 2.23 & 9.75 & 16.4$\pm$2.59 & -20.68 & 2.00 & 13.9 & 13.3$\pm$3.37 & -9.56 & 1.11 & 1.24 \\
\tableline
\multirow{8}{*}{A12} & \multirow{2}{*}{0.1} & \multirow{2}{*}{9.16} & LTE & 1.66 & 3.77 & 5.29$\pm$1.45 & -9.33 & 1.22 & 1.93 & 1.22$\pm$1.73 & -1.37 & 1.37 & 4.33 \\
   & & & NLTE & 1.50 & 2.83 & 3.84$\pm$1.36 & -6.73 & 1.15 & 1.36 & 1.35$\pm$1.72 & -1.52 & 1.30 & 2.85 \\ 
   \cline{2-14}
   & \multirow{2}{*}{5} & \multirow{2}{*}{6.03} & LTE & 2.01 & 4.36 & 5.77$\pm$1.24 & -13.33 & 1.31 & 2.04 & 2.12$\pm$1.34 & -3.19 & 1.54 & 2.72 \\
   & & & NLTE & 1.87 & 3.74 & 5.05$\pm$1.18 & -11.69 & 1.34 & 2.28 & 2.15$\pm$1.34 & -3.26 & 1.39 & 2.35 \\  
   \cline{2-14}
   & \multirow{2}{*}{10} & \multirow{2}{*}{4.28} & LTE & 1.71 & 2.39 & 3.43$\pm$0.88 & -10.19 & 1.46 & 2.30 & 2.15$\pm$1.04 & -4.32 & 1.17 & 1.60 \\
   & & & NLTE & 1.82 & 2.76 & 4.25$\pm$0.93 & -12.63 & 1.41 & 2.03 & 2.01$\pm$1.03 & -4.04 & 1.29 & 2.11 \\ 
   \cline{2-14}
   & \multirow{2}{*}{20} & \multirow{2}{*}{2.78} & LTE & 1.33 & 0.75 & 0.96$\pm$0.51 & -4.21 & 1.35 & 1.23 & 1.30$\pm$0.72 & -3.71 & 0.98 & 0.74 \\
   & & & NLTE & 1.34 & 0.79 & 0.63$\pm$0.44 & -2.75 & 1.39 & 1.35 & 1.32$\pm$0.71 & -3.81 & 0.97 & 0.48 \\
\tablebreak
\multirow{8}{*}{A13} & \multirow{2}{*}{0.1} & \multirow{2}{*}{16.2} & LTE & 2.11 & 9.79 & 15.0$\pm$2.70 & -16.97 & 1.62 & 9.46 & 9.63$\pm$3.33 & -6.35 & 1.30 & 1.56 \\
   & & & NLTE & 2.05 & 9.29 & 14.1$\pm$2.64 & -15.86 & 1.56 & 8.49 & 7.75$\pm$3.23 & -5.12 & 1.31 & 1.81 \\ 
   \cline{2-14}
   & \multirow{2}{*}{5} & \multirow{2}{*}{11.6} & LTE & 2.32 & 9.85 & 15.3$\pm$2.46 & -20.52 & 1.79 & 9.61 & 10.1$\pm$2.81 & -8.26 & 1.30 & 1.52 \\
   & & & NLTE & 2.23 & 9.19 & 14.6$\pm$2.40 & -19.58 & 1.81 & 9.95 & 9.08$\pm$2.79 & -7.41 & 1.23 & 1.61 \\  
   \cline{2-14}
   & \multirow{2}{*}{10} & \multirow{2}{*}{7.80} & LTE & 1.86 & 4.85 & 7.46$\pm$1.58 & -13.16 & 1.97 & 8.85 & 7.58$\pm$2.14 & -8.34 & 0.94 & 0.98 \\
   & & & NLTE & 2.07 & 6.06 & 9.04$\pm$1.69 & -15.99 & 1.98 & 8.81 & 7.72$\pm$2.13 & -8.56 & 1.05 & 1.17 \\ 
   \cline{2-14}
   & \multirow{2}{*}{20} & \multirow{2}{*}{4.95} & LTE & 1.43 & 1.77 & 1.87$\pm$0.93 & -4.53 & 1.57 & 3.55 & 2.60$\pm$1.30 & -4.20 & 0.91 & 0.72 \\
   & & & NLTE & 1.46 & 1.93 & 2.38$\pm$0.96 & -5.73 & 1.59 & 3.61 & 2.92$\pm$1.31 & -4.74 & 0.92 & 0.82 \\
\tableline
\multirow{8}{*}{A14} & \multirow{2}{*}{0.1} & \multirow{2}{*}{24.0} & LTE & 3.12 & 22.4 & 39.2$\pm$4.60 & -37.10 & 2.42 & 28.6 & 29.5$\pm$5.47 & -14.69 & 1.29 & 1.33 \\
   & & & NLTE & 2.97 & 20.8 & 35.2$\pm$4.35 & -33.17 & 2.29 & 25.8 & 25.2$\pm5.19$ & -12.64 & 1.29 & 1.40 \\ 
   \cline{2-14}
   & \multirow{2}{*}{5} & \multirow{2}{*}{18.9} & LTE & 3.03 & 20.6 & 36.0$\pm$4.29 & -35.54 & 2.61 & 2.94 & 28.1$\pm$5.15 & -15.31 & 1.16 & 1.28 \\
   & & & NLTE & 3.24 & 22.6 & 37.8$\pm$4.45 & -37.45 & 2.43 & 26.4 & 27.8$\pm$5.06 & -15.07 & 1.33 & 1.36 \\  
   \cline{2-14}
   & \multirow{2}{*}{10} & \multirow{2}{*}{13.1} & LTE & 2.63 & 13.3 & 23.1$\pm$3.07 & -28.26 & 2.74 & 24.8 & 25.1$\pm$4.19 & -17.62 & 0.96 & 0.92 \\
   & & & NLTE & 2.69 & 13.8 & 22.5$\pm$3.07 & -27.54 & 2.68 & 24.0 & 25.1$\pm$4.17 & -17.55 & 1.00 & 0.90 \\ 
   \cline{2-14}
   & \multirow{2}{*}{20} & \multirow{2}{*}{7.93} & LTE & 1.71 & 4.37 & 5.66$\pm$1.58 & -9.18 & 1.81 & 8.00 & 6.78$\pm$2.24 & -6.83 & 0.95 & 0.83 \\
   & & & NLTE & 1.81 & 4.92 & 6.36$\pm$1.62 & -10.42 & 1.73 & 7.24 & 6.53$\pm$2.22 & -6.56 & 1.05 & 0.97 \\
\tableline
\multirow{8}{*}{A15} & \multirow{2}{*}{0.1} & \multirow{2}{*}{34.7} & LTE & 5.00 & 45.4 & 85.4$\pm$7.68 & -75.1 & 3.97 & 71.8 & 72.5$\pm$9.42 & -30.00 & 1.26 & 1.18 \\
   & & & NLTE & 4.90 & 44.2 & 83.7$\pm$7.56 & -73.48 & 3.89 & 69.5 & 67.9$\pm$9.14 & -28.18 & 1.26 & 1.23 \\ 
   \cline{2-14}
   & \multirow{2}{*}{5} & \multirow{2}{*}{30.3} & LTE & 4.31 & 39.6 & 74.9$\pm$7.11 & -62.21 & 3.58 & 64.1 & 68.9$\pm$9.05 & -27.72 & 1.20 & 1.09 \\
   & & & NLTE & 4.53 & 43.0 & 80.9$\pm$7.55 & -66.33 & 3.55 & 63.6 & 65.7$\pm$8.94 & -26.31 & 1.28 & 1.23 \\  
   \cline{2-14}
   & \multirow{2}{*}{10} & \multirow{2}{*}{23.2} & LTE & 2.81 & 19.5 & 39.7$\pm$45.8 & -36.08 & 3.67 & 59.1 & 65.2$\pm$8.26 & -29.49 & 0.77 & 0.61 \\
   & & & NLTE & 4.40 & 37.6 & 69.3$\pm$6.62 & -62.44 & 3.58 & 56.8 & 62.4$\pm$8.07 & -28.35 & 1.23 & 1.11 \\ 
   \cline{2-14}
   & \multirow{2}{*}{20} & \multirow{2}{*}{13.7} & LTE & 1.75 & 6.46 & 9.72$\pm$2.31 & -11.21 & 2.27 & 20.1 & 20.6$\pm$4.17 & -12.99 & 0.77 & 0.47 \\
   & & & NLTE & 2.50 & 13.1 & 20.6$\pm$3.07 & -23.57 & 2.21 & 19.2 & 21.1$\pm$4.17 & -13.26 & 1.13 & 0.98
\enddata
\tablecomments{The amplitudes have suffices CN and CS for continuum normalized and continuum subtracted spectrum, respectively. The continuum subtracted spectrum has a unit of \ergpsqcmpspang . The line integrated flux in the units of \ergpsqcmps \ is estimated using the line$\_$flux routine from the Specutils Python package.}
\end{deluxetable*}
\end{longrotatetable}

\begin{longrotatetable}
\begin{deluxetable*}{cccccccccccccc}
\tablecaption{Hydrogen Line Properties for B Series Models. \label{tab:B_Series}}
\tabletypesize{\footnotesize}
\tablewidth{0pt}
\tablehead{
\colhead{Model} & \colhead{day} & \colhead{$\rm L_{bol}$} & \colhead{H-Opacity} & \colhead{A(\halpha)$_{CN}$} & \colhead{A(\halpha)$_{CS}$} & \colhead{F(\halpha)} & \colhead{EW(\halpha)} & \colhead{A(\hbeta)$_{CN}$} & \colhead{A(\halpha)$_{CS}$} & \colhead{F(\hbeta)} & \colhead{EW(\hbeta)} & \colhead{A(\halpha)$_{CN}$/} & \colhead{F(\halpha)/} \\ [-.2cm]
\colhead{} & \colhead{} & \colhead{$\rm (\times10^{42})$} & \colhead{} & \colhead{} & \colhead{$\rm (\times10^{-2})$} & \colhead{$\rm (\times10^{-1})$} & \colhead{} & \colhead{} & \colhead{$\rm (\times10^{-2})$} & \colhead{$\rm (\times10^{-1})$} & \colhead{} & \colhead{A(\halpha)$_{CN}$} & \colhead{F(\hbeta)}
}
\decimalcolnumbers
\startdata
\multirow{8}{*}{B4} & \multirow{2}{*}{0.1} & \multirow{2}{*}{2.56} & LTE & 3.51 & 4.22 & 6.05$\pm$0.76 & -35.93 & 2.73 & 4.34 & 4.37$\pm$0.74 & -17.40 & 1.29 & 1.38 \\
   & & & NLTE & 3.67 & 4.51 & 6.54$\pm$0.80 & -38.74 & 2.78 & 4.44 & 4.23$\pm$0.73 & -16.93 & 1.32 & 1.55 \\ 
   \cline{2-14}
   & \multirow{2}{*}{5} & \multirow{2}{*}{2.20} & LTE & 1.70 & 1.17 & 1.60$\pm$0.43 & -9.50 & 2.01 & 2.53 & 2.28$\pm$0.60 & -9.11 & 0.84 & 0.70 \\
   & & & NLTE & 1.97 & 1.62 & 2.29$\pm$4.78 & -13.70 & 2.03 & 2.55 & 2.52$\pm$0.60 & -10.19 & 0.97 & 0.91 \\  
   \cline{2-14}
   & \multirow{2}{*}{10} & \multirow{2}{*}{1.65} & LTE & 1.49 & 0.67 & 7.31$\pm$3.13 & -5.41 & 1.66 & 1.32 & 1.33$\pm$4.42 & -6.71 & 0.90 & 0.55 \\
   & & & NLTE & 1.58 & 0.78 & 0.92$\pm$0.32 & -6.86 & 1.69 & 1.36 & 1.20$\pm$0.44 & -6.10 & 0.94 & 0.77 \\ 
   \cline{2-14}
   & \multirow{2}{*}{20} & \multirow{2}{*}{0.74} & LTE & 1.46 & 2.97 & 4.22$\pm$1.52 & -6.53 & 1.34 & 3.17 & 2.10$\pm$1.87 & -2.24 & 1.09 & 2.01 \\
   & & & NLTE & 1.48 & 3.04 & 4.15$\pm$1.51 & -6.50 & 1.30 & 2.79 & 2.00$\pm$1.86 & -2.14 & 1.14 & 2.08 \\
\tableline
\multirow{8}{*}{B5} & \multirow{2}{*}{0.1} & \multirow{2}{*}{4.25} & LTE & 3.69 & 7.01 & 1.08$\pm$0.126 & -41.60 & 4.33 & 1.41 & 1.41$\pm$0.176 & -33.31 & 0.85 & 0.77 \\
   & & & NLTE & 4.07 & 8.03 & 1.26 $\pm$ 0.138 & -48.01 & 4.26 & 1.38 & 1.32 $\pm$ 0.172 & -31.17 & 0.96 & 0.95 \\ 
   \cline{2-14}
   & \multirow{2}{*}{5} & \multirow{2}{*}{3.77} & LTE & 2.37 & 3.63 & 5.08$\pm$0.87 & -19.20 & 2.42 & 6.06 & 5.06$\pm$1.13 & -11.84 & 0.98 & 1.00 \\
   & & & NLTE & 2.58 & 4.18 & 5.61$\pm$9.15 & -21.23 & 2.48 & 6.24 & 5.25$\pm$1.13 & -12.43 & 1.04 & 1.07 \\  
   \cline{2-14}
   & \multirow{2}{*}{10} & \multirow{2}{*}{2.53} & LTE & 2.39 & 2.78 & 3.72$\pm$6.50 & -18.60 & 1.83 & 2.63 & 1.87$\pm$7.07 & -5.93 & 1.30 & 1.99 \\
   & & & NLTE & 2.39 & 2.78 & 3.72$\pm$6.50 & -18.60 & 1.83 & 2.63 & 1.87$\pm$7.07 & -5.93 & 1.30 & 1.99 \\ 
   \cline{2-14}
   & \multirow{2}{*}{20} & \multirow{2}{*}{1.10} & LTE & 1.65 & 6.31 & 8.16$\pm$2.44 & -8.34 & 1.31 & 4.47 & 3.41$\pm$2.87 & -2.37 & 1.26 & 2.39 \\
   & & & NLTE & 1.69 & 6.76 & 7.49$\pm$2.43 & -7.66 & 1.33 & 4.73 & 4.17$\pm$2.89 & -2.93 & 1.27 & 1.80 \\
\tableline
\multirow{8}{*}{B6} & \multirow{2}{*}{0.1} & \multirow{2}{*}{6.29} & LTE & 5.09 & 1.39 & 2.28$\pm$2.23 & -66.88 & 4.40 & 2.10 & 2.19$\pm$2.65 & -35.58 & 1.16 & 1.04 \\
   & & & NLTE & 5.71 & 1.59 & 2.56$\pm$2.45 & -75.75 & 4.47 & 2.15 & 2.27$\pm$2.69 & -36.60 & 1.28 & 1.13 \\ 
   \cline{2-14}
   & \multirow{2}{*}{5} & \multirow{2}{*}{5.59} & LTE & 3.93 & 1.01 & 1.56$\pm$1.75 & -45.52 & 2.91 & 1.15 & 1.19$\pm$1.88 & -19.63 & 1.35 & 1.32 \\
   & & & NLTE & 4.03 & 1.05 & 1.65$\pm$1.81 & -47.67 & 2.96 & 1.19 & 1.16$\pm$1.89 & -19.10 & 1.36 & 1.42 \\  
   \cline{2-14}
   & \multirow{2}{*}{10} & \multirow{2}{*}{4.01} & LTE & 3.42 & 6.65 & 9.98$\pm$1.23 & -36.26 & 2.17 & 5.66 & 5.66$\pm$1.24 & -11.71 & 1.57 & 1.76 \\
   & & & NLTE & 3.47 & 6.84 & 1.03$\pm$1.26 & -36.97 & 2.25 & 5.97 & 5.47$\pm$1.23 & -11.43 & 1.54 & 1.88 \\ 
   \cline{2-14}
   & \multirow{2}{*}{20} & \multirow{2}{*}{1.83} & LTE & 1.88 & 1.44 & 1.84$\pm$4.43 & -11.26 & 1.40 & 0.99 & 0.57$\pm$0.49 & -2.36 & 1.35 & 3.20 \\
   & & & NLTE & 1.95 & 1.54 & 2.32$\pm$0.47 & -14.40 & 1.37 & 0.92 & 0.41$\pm$0.49 & -1.68 & 1.42 & 5.66 \\
\tablebreak
\multirow{8}{*}{B7} & \multirow{2}{*}{0.1} & \multirow{2}{*}{8.98} & LTE & 7.19 & 2.55 & 4.39$\pm$3.77 & -106.57 & 4.35 & 2.87 & 2.92$\pm$3.63 & -34.14 & 1.65 & 1.50 \\
   & & & NLTE & 7.55 & 2.76 & 4.66$\pm$4.01 & -110.65 & 4.47 & 2.96 & 2.91$\pm$3.68 & -34.01 & 1.69 & 1.60 \\ 
   \cline{2-14}
   & \multirow{2}{*}{5} & \multirow{2}{*}{8.14} & LTE & 5.20 & 1.80 & 3.13$\pm$2.92 & -72.91 & 3.32 & 1.95 & 1.74$\pm$2.80 & -20.67 & 1.56 & 1.80 \\
   & & & NLTE & 5.59 & 2.00 & 3.34$\pm$3.12 & -76.82 & 3.39 & 2.04 & 1.82$\pm$2.88 & -21.37 & 1.65 & 1.84 \\  
   \cline{2-14}
   & \multirow{2}{*}{10} & \multirow{2}{*}{6.44} & LTE & 4.33 & 1.25 & 2.06$\pm$2.13 & -54.61 & 2.70 & 1.22 & 1.18$\pm$2.09 & -16.47 & 1.60 & 1.75 \\
   & & & NLTE & 4.77 & 1.43 & 2.36$\pm$2.35 & -62.26 & 2.70 & 1.24 & 1.18$\pm$2.12 & -16.20 & 1.77 & 2.00 \\ 
   \cline{2-14}
   & \multirow{2}{*}{20} & \multirow{2}{*}{3.62} & LTE & 2.56 & 4.22 & 6.14$\pm$9.49 & -22.68 & 1.59 & 2.80 & 2.53$\pm$1.03 & -5.31 & 1.61 & 2.42 \\
   & & & NLTE & 2.67 & 4.50 & 6.74 $\pm$ 9.88 & -25.01 & 1.60 & 2.86 & 2.47 $\pm$ 1.04 & -5.14 & 1.68 & 2.72 \\
\tableline
\multirow{8}{*}{B8} & \multirow{2}{*}{0.1} & \multirow{2}{*}{5.11} & LTE & 2.83 & 0.06 & 0.89$\pm$0.13 & -26.40 & 1.59 & 0.30 & 2.69$\pm$1.10 & -5.27 & 1.79 & 3.31 \\
   & & & NLTE & 2.64 & 5.45 & 7.44$\pm$1.17 & -22.33 & 1.52 & 2.65 & 2.81$\pm$1.09 & -5.53 & 1.73 & 2.65 \\ 
   \cline{2-14}
   & \multirow{2}{*}{5} & \multirow{2}{*}{3.61} & LTE & 2.20 & 3.20 & 4.78$\pm$8.37 & -17.87 & 1.68 & 2.80 & 3.29$\pm$9.32 & -8.02 & 1.31 & 1.45 \\
   & & & NLTE & 2.28 & 3.39 & 4.93$\pm$0.85 & -18.54 & 1.70 & 2.89 & 3.31$\pm$0.94 & -8.01 & 1.34 & 1.49 \\  
   \cline{2-14}
   & \multirow{2}{*}{10} & \multirow{2}{*}{2.68} & LTE & 1.63 & 1.33 & 1.92$\pm$0.53 & -9.18 & 1.73 & 2.35 & 2.21$\pm$0.72 & -6.85 & 0.95 & 0.87 \\
   & & & NLTE & 1.77 & 1.60 & 2.28$\pm$0.55 & -10.95 & 1.69 & 2.23 & 1.86$\pm$7.11 & -5.73 & 1.05 & 1.22 \\ 
   \cline{2-14}
   & \multirow{2}{*}{20} & \multirow{2}{*}{1.55} & LTE & 1.48 & 0.62 & 0.98$\pm$0.31 & -7.57 & 1.32 & 0.66 & 0.51$\pm$0.41 & -2.51 & 1.11 & 1.93 \\
   & & & NLTE & 1.48 & 0.63 & 0.91$\pm$0.31 & -6.96 & 1.38 & 0.76 & 0.25$\pm$0.40 & -1.23 & 1.08 & 3.67 \\
\tableline
\multirow{8}{*}{B9} & \multirow{2}{*}{0.1} & \multirow{2}{*}{8.82} & LTE & 4.12 & 15.50 & 24.50$\pm$26.60 & -49.29 & 2.80 & 15.20 & 15.90$\pm$25.30 & -18.80 & 1.47 & 1.54 \\
   & & & NLTE & 4.23 & 16.10 & 24.80$\pm$27.10 & -49.72 & 2.78 & 14.90 & 14.00$\pm$24.50 & -16.67 & 1.52 & 1.77 \\ 
   \cline{2-14}
   & \multirow{2}{*}{5} & \multirow{2}{*}{7.03} & LTE &  2.52 & 7.18 & 11.50$\pm$16.70 & -24.34 & 2.61 & 12.20 & 11.10$\pm$21.20 & -14.62 & 0.97 & 1.03 \\
   & & & NLTE &2.76 & 8.33 & 12.70$\pm$17.80 & -26.79 & 2.60 & 12.00 & 10.80$\pm$20.90 & -14.29 & 1.06 & 1.18 \\  
   \cline{2-14}
   & \multirow{2}{*}{10} & \multirow{2}{*}{5.15} & LTE & 1.84 & 3.26 & 4.46$\pm$1.05 & -26.79 & 2.27 & 7.82 & 6.46$\pm$1.56 & -14.29 & 0.81 & 0.69 \\
   & & & NLTE & 2.00 & 3.87 & 5.23$\pm$1.10 & -11.50 & 2.22 & 7.47 & 6.37$\pm$1.54 & -10.50 & 0.90 & 0.82 \\ 
   \cline{2-14}
   & \multirow{2}{*}{20} & \multirow{2}{*}{2.87} & LTE & 1.72 & 1.79 & 1.64$\pm$6.11 & -13.55 & 1.45 & 1.69 & 1.45$\pm$7.73 & -10.36 & 1.18 & 1.13 \\
   & & & NLTE & 1.76 & 1.87 & 1.96$\pm$6.22 & -6.55 & 1.48 & 1.77 & 1.22$\pm$7.60 & -3.89 & 1.19 & 1.61 \\
\tablebreak
\multirow{8}{*}{B10} & \multirow{2}{*}{0.1} & \multirow{2}{*}{13.10} & LTE & 5.41 & 2.74 & 4.60$\pm$4.34 & -73.82 & 4.67 & 4.27 & 4.38$\pm$5.18 & -37.53 & 1.16 & 1.05 \\
   & & & NLTE & 5.52 & 2.84 & 4.96$\pm$4.54 & -78.99 & 4.68 & 4.27 & 4.04$\pm$5.09 & -34.77 & 1.18 & 1.23 \\ 
   \cline{2-14}
   & \multirow{2}{*}{5} & \multirow{2}{*}{17.00} & LTE & 3.54 & 1.55 & 2.71$\pm$2.95 & -44.40 & 3.69 & 2.95 & 3.03$\pm$4.04 & -27.60 & 0.96 & 0.90 \\
   & & & NLTE & 3.60 & 1.60 & 2.72$\pm$2.99 & -44.37 & 3.58 & 2.83 & 2.92$\pm$3.94 & -26.60 & 1.00 & 0.93 \\  
   \cline{2-14}
   & \multirow{2}{*}{10} & \multirow{2}{*}{8.26} & LTE & 2.39 & 0.75 & 1.19$\pm$0.18 & -22.06 & 2.70 & 1.59 & 1.58$\pm$0.27 & -16.90 & 0.89 & 0.75 \\
   & & & NLTE & 2.62 & 0.88 & 1.39$\pm$0.20 & -25.52 & 2.75 & 1.62 & 1.56$\pm$0.27 & -16.82 & 0.95 & 0.89 \\ 
   \cline{2-14}
   & \multirow{2}{*}{20} & \multirow{2}{*}{4.76} & LTE & 2.14 & 0.45 & 5.33$\pm$1.15 & -13.58 & 1.62 & 0.38 & 2.41$\pm$1.31 & -3.88 & 1.32 & 2.21 \\
   & & & NLTE & 2.24 & 4.84 & 6.28$\pm$1.20 & -16.11 & 1.63 & 3.88 & 2.63$\pm$1.30 & -4.27 & 1.38 & 2.39 \\
\tableline
\multirow{8}{*}{B11} & \multirow{2}{*}{0.1} & \multirow{2}{*}{18.10} & LTE & 3.44 & 1.41 & 3.71 $\pm$ 3.62 & -53.12 & 5.50 & 6.77 & 7.30 $\pm$ 7.89 & -48.46 & 0.63 & 0.51 \\
   & & & NLTE & 7.40 & 4.47 & 8.09 $\pm$ 6.70 & -115.40 & 5.48 & 6.71 & 6.87 $\pm$ 7.75 & -45.81 & 1.35 & 1.18 \\ 
   \cline{2-14}
   & \multirow{2}{*}{5} & \multirow{2}{*}{15.70} & LTE & 2.49 & 0.91 & 2.33$\pm$0.29 & -31.79 & 4.05 & 4.48 & 4.80$\pm$5.96 & -32.58 & 0.62 & 0.48 \\
   & & & NLTE & 4.76 & 2.81 & 5.13$\pm$4.78 & -68.45 & 4.11 & 4.57 & 4.76$\pm$5.97 & -32.39 & 1.16 & 1.08 \\  
   \cline{2-14}
   & \multirow{2}{*}{10} & \multirow{2}{*}{12.90} & LTE & 3.13 & 1.50 & 2.78$\pm$3.13 & -39.34 & 2.85 & 2.47 & 2.65$\pm$4.14 & -19.79 & 1.10 & 1.05 \\
   & & & NLTE & 3.45 & 1.74 & 3.00$\pm$3.34 & -42.32 & 2.82 & 2.48 & 2.65$\pm$4.19 & -19.49 & 1.22 & 1.13 \\ 
   \cline{2-14}
   & \multirow{2}{*}{20} & \multirow{2}{*}{8.29} & LTE & 2.53 & 8.86 & 1.29 $\pm$ 2.01 & -22.12 & 1.86 & 8.85 & 5.89 $\pm$ 2.31 & -5.73 & 1.36 & 2.18 \\
   & & & NLTE & 2.86 & 1.08 & 1.63$\pm$2.25 & -28.07 & 1.93 & 9.60 & 6.82$\pm$2.37 & -6.60 & 1.48 & 2.39 \\
\tableline
\multirow{8}{*}{B12} & \multirow{2}{*}{0.1} & \multirow{2}{*}{10.80} & LTE & 1.44 & 2.97 & 3.89$\pm$1.57 & -5.71 & 1.21 & 2.21 & 2.81$\pm$2.14 & -2.61 & 1.19 & 1.38 \\
   & & & NLTE &  1.45 & 3.05 & 4.54$\pm$1.60 & -6.69 & 1.07 & 8.05 & 5.13$\pm$2.03 & -0.48 & 1.35 & 8.85 \\ 
   \cline{2-14}
   & \multirow{2}{*}{5} & \multirow{2}{*}{7.13} & LTE & 1.63 & 3.26 & 4.76$\pm$1.31 & -9.16 & 1.21 & 1.65 & 2.31$\pm$1.59 & -2.91 & 1.35 & 2.06 \\
   & & & NLTE & 1.52 & 2.72 & 3.57$\pm$1.25 & -6.79 & 1.18 & 1.46 & 1.76$\pm$1.56 & -2.23 & 1.28 & 2.03 \\  
   \cline{2-14}
   & \multirow{2}{*}{10} & \multirow{2}{*}{4.73} & LTE & 1.77 & 2.86 & 4.41$\pm$1.00 & -11.89 & 1.25 & 1.39 & 1.50$\pm$1.09 & -2.74 & 1.41 & 2.94 \\
   & & & NLTE & 1.72 & 2.68 & 3.93$\pm$0.98 & -10.57 & 1.26 & 1.45 & 1.89$\pm$1.11 & -3.43 & 1.36 & 2.08 \\ 
   \cline{2-14}
   & \multirow{2}{*}{20} & \multirow{2}{*}{3.26} & LTE & 1.32 & 0.87 & 0.09$\pm$0.06 & -3.47 & 1.30 & 1.19 & 0.10$\pm$0.08 & -2.42 & 1.01 & 1.00 \\
   & & & NLTE & 1.45 & 1.24 & 0.20$\pm$0.07 & -7.09 & 1.26 & 1.00 & 0.09$\pm$0.08 & -2.38 & 1.16 & 2.09 \\
\tablebreak
\multirow{8}{*}{B13} & \multirow{2}{*}{0.1} & \multirow{2}{*}{18.40} & LTE & 1.99 & 9.56 & 0.15$\pm$0.03 & -15.12 & 1.44 & 7.48 & 0.70$\pm$0.36 & -4.06 & 1.38 & 2.11 \\
   & & & NLTE & 1.95 & 9.24 & 0.15$\pm$0.03 & -13.27 & 1.43 & 7.32 & 0.82$\pm$0.36 & -4.77 & 1.37 & 1.58 \\ 
   \cline{2-14}
   & \multirow{2}{*}{5} & \multirow{2}{*}{13.60} & LTE & 2.11 & 9.67 & 0.15$\pm$0.03 & -17.25 & 1.57 & 8.35 & 0.76$\pm$0.31 & -5.25 & 1.34 & 1.97 \\
   & & & NLTE & 1.92 & 8.05 & 0.15$\pm$0.03 & -14.49 & 1.45 & 6.56 & 0.66$\pm$0.30 & -4.54 & 1.33 & 1.92 \\  
   \cline{2-14}
   & \multirow{2}{*}{10} & \multirow{2}{*}{8.84} & LTE & 2.01 & 6.46 & 1.05$\pm$0.19 & -16.51 & 1.79 & 8.27 & 0.63$\pm$0.23 & -6.06 & 1.12 & 1.79 \\
   & & & NLTE & 2.12 & 7.17 & 1.12$\pm$0.20 & -17.44 & 1.82 & 8.50 & 0.80$\pm$0.24 & -7.73 & 1.16 & 1.82 \\ 
   \cline{2-14}
   & \multirow{2}{*}{20} & \multirow{2}{*}{5.57} & LTE & 1.32 & 1.47 & 1.45$\pm$1.00 & -3.12 & 1.60 & 4.25 & 0.45$\pm$0.35 & -4.56 & 0.82 & 0.35 \\
   & & & NLTE & 1.46 & 2.16 & 2.63$\pm$1.08 & -5.62 & 1.52 & 3.69 & 0.92$\pm$0.59 & -4.04 & 0.96 & 0.59 \\
\tableline
\multirow{8}{*}{B14} & \multirow{2}{*}{0.1} & \multirow{2}{*}{28.40} & LTE & 2.86 & 21.7 & 38.3$\pm$47.2 & -32.86 & 2.20 & 27.4 & 26.8$\pm$58.1 & -11.73 & 1.30 & 1.43 \\
   & & & NLTE & 2.76 & 20.7 & 33.4$\pm$44.9 & -28.40 & 2.13 & 25.9 & 26.8$\pm$57.7 & -11.72 & 1.30 & 1.24 \\ 
   \cline{2-14}
   & \multirow{2}{*}{5} & \multirow{2}{*}{22.7} & LTE & 2.78 & 20.7 & 36.3$\pm$46.0 & -31.14 & 2.09 & 23.8 & 25.8$\pm$54.8 & -11.84 & 1.33 & 1.41 \\
   & & & NLTE & 2.55 & 18.3 & 32.7$\pm$43.7 & -27.74 & 1.93 & 20.2 & 19.3$\pm$51.5 & -8.88 & 1.32 & 1.69 \\  
   \cline{2-14}
   & \multirow{2}{*}{10} & \multirow{2}{*}{15.40} & LTE & 2.80 & 16.9 & 29.5$\pm$37.2 & -31.46 & 2.51 & 25.4 & 26.4$\pm$46.9 & -15.73 & 1.12 & 1.12 \\
   & & & NLTE & 2.84 & 17.3 & 28.8$\pm$37.2 & -30.57 & 2.36 & 22.9 & 23.4$\pm$45.0 & -13.90 & 1.20 & 1.23 \\ 
   \cline{2-14}
   & \multirow{2}{*}{20} & \multirow{2}{*}{8.83} & LTE & 1.68 & 4.59 & 5.58$\pm$1.70 & -8.22 & 1.80 & 8.94 & 7.79$\pm$2.53 & -6.94 & 0.93 & 0.72 \\
   & & & NLTE & 1.74 & 5.08 & 7.66$\pm$1.82 & -11.21 & 1.74 & 8.34 & 7.75$\pm$2.52 & -6.89 & 1.00 & 0.99 \\
\tableline
\multirow{8}{*}{B15} & \multirow{2}{*}{0.1} & \multirow{2}{*}{41.00} & LTE & 4.44 & 4.32 & 7.97$\pm$7.58 & -63.24 & 3.42 & 6.67 & 6.97$\pm$9.60 & -25.29 & 1.30 & 1.14 \\
   & & & NLTE & 4.26 & 4.11 & 8.09$\pm$7.55 & -63.83 & 3.23 & 6.16 & 6.05$\pm$9.10 & -21.93 & 1.32 & 1.34 \\ 
   \cline{2-14}
   & \multirow{2}{*}{5} & \multirow{2}{*}{35.40} & LTE & 3.96 & 4.00 & 7.39$\pm$7.37 & -54.59 & 3.16 & 6.16 & 6.45$\pm$9.39 & -22.62 & 1.25 & 1.15 \\
   & & & NLTE & 3.96 & 4.01 & 7.58$\pm$7.45 & -55.77 & 3.10 & 6.02 & 6.25$\pm$9.28 & -21.82 & 1.28 & 1.21 \\  
   \cline{2-14}
   & \multirow{2}{*}{10} & \multirow{2}{*}{26.60} & LTE & 4.07 & 3.78 & 7.17$\pm$6.95 & -58.17 & 3.39 & 5.97 & 6.74$\pm$8.84 & -26.94 & 1.20 & 1.06 \\
   & & & NLTE & 4.20 & 3.97 & 7.37$\pm$7.15 & -59.37 & 3.26 & 5.69 & 6.00$\pm$8.52 & -23.83 & 1.29 & 1.23 \\ 
   \cline{2-14}
   & \multirow{2}{*}{20} & \multirow{2}{*}{15.10} & LTE & 1.72 & 6.88 & 1.13$\pm$2.56 & -11.83 & 2.32 & 2.32 & 2.52$\pm$4.74 & -14.28 & 0.74 & 0.45 \\
   & & & NLTE & 2.58 & 1.52 & 2.57$\pm$3.53 & -26.69 & 2.33 & 2.33 & 2.56$\pm$4.75 & -14.54 & 1.11 & 1.00
\enddata
\tablecomments{The amplitudes have suffices CN and CS for continuum normalized and continuum subtracted spectrum, respectively. The continuum subtracted spectrum has a unit of \ergpsqcmpspang . The line integrated flux in the units of \ergpsqcmps \ is estimated using the line$\_$flux routine from the Specutils Python package.}
\end{deluxetable*}
\end{longrotatetable}

%\begin{equation}\label{eqn:norm_B}
%,
%\end{equation}

%\begin{eqnarray}\label{eqn:IMC_ss}
% \nonumber \\
% \nonumber \\
%
%\;\;\;\;. \quad
%\end{eqnarray}

%\begin{eqnarray}
%d_B = 
%\left\{
%\begin{array}{ll}
% \\
%\quad \quad {\rm if} \ \mu < -\sqrt{1-(x_i/x)^2} \\
% \\
%\quad \quad {\rm otherwise}
%\end{array}
%\right.
%\end{eqnarray}

%\begin{figure*}[ht!]
%\begin{center}
%\includegraphics[width=0.9\textwidth]{.png}
%\caption{The . 
%\label{fig:w7_prof}}
%\end{center}
%\end{figure*}

%\begin{longrotatetable}
%\begin{deluxetable*}{lllrrrrrrll}
%\tablecaption{\label{}}
%\tablewidth{700pt}
%\tabletypesize{\scriptsize}
%\tablehead{
%\colhead{()} & \colhead{()} & \colhead{} & \colhead{}
%} 
%\startdata
% \\
%\enddata
%\end{deluxetable*}
%\end{longrotatetable}

%% Appendix material should be preceded with a single \appendix command.
%% There should be a \section command for each appendix. Mark appendix
%% subsections with the same markup you use in the main body of the paper.

%% Each Appendix (indicated with \section) will be lettered A, B, C, etc.
%% The equation counter will reset when it encounters the \appendix
%% command and will number appendix equations (A1), (A2), etc. The
%% Figure and Table counter will not reset.

%% For this sample we use BibTeX plus aasjournals.bst to generate the
%% the bibliography. The sample631.bib file was populated from ADS. To
%% get the citations to show in the compiled file do the following:
%%
%% pdflatex sample631.tex
%% bibtext sample631
%% pdflatex sample631.tex
%% pdflatex sample631.tex

\bibliography{superlite.bib}{}
\bibliographystyle{aasjournal}

%% This command is needed to show the entire author+affiliation list when
%% the collaboration and author truncation commands are used.  It has to
%% go at the end of the manuscript.
%\allauthors

%% Include this line if you are using the \added, \replaced, \deleted
%% commands to see a summary list of all changes at the end of the article.
%\listofchanges

\end{document}